\documentclass[aps,prx,twocolumn,amsmath,amssymb,10pt]{revtex4-1}
\usepackage{graphicx}
\usepackage{subfigure}
\usepackage[colorlinks, linkcolor=blue]{hyperref}
\usepackage[usenames,dvipsnames]{color}
\usepackage{verbatim}
\usepackage{txfonts}
\usepackage{ulem}
\def \k {{\bf k}}

\def \e {\varepsilon}

\def \Q{{\bf Q}}

\def \D{\Delta}

\def \ol{\overline}

\def \beq {\begin{eqnarray}}
\def \eeq {\end{eqnarray}}

\def \D{\Delta}
\def \1{\bf 1}
\def \2{\bf 2}

\def \tn{\textnormal}

\def \e{\varepsilon}

\def \ol{\overline}
\newcommand{\ii}{\mathrm{i}}
\newcommand{\dd}{\mathrm{d}}
\renewcommand{\vec}[1]{\boldsymbol{#1}}
\renewcommand{\Re}{\mathrm{Re}\,}
\renewcommand{\Im}{\mathrm{Im}\,}
\newcommand{\ket}[1]{\left|#1\right\rangle}
\newcommand{\bra}[1]{\left\langle#1\right|}

\newcommand{\etal}{\textit{et al.\,}}
\begin{document}
\title{Connecting high-field quantum oscillations to\\ zero-field electron spectral functions in the underdoped cuprates}
\author{Andrea Allais}
\affiliation{Department of Physics, Harvard University, Cambridge Massachusetts
02138, USA.}
\author{Debanjan Chowdhury}
\affiliation{Department of Physics, Harvard University, Cambridge Massachusetts
02138, USA.}
\author{Subir Sachdev}
\affiliation{Department of Physics, Harvard University, Cambridge Massachusetts
02138, USA,}
\affiliation{Perimeter Institute for Theoretical Physics, Waterloo, Ontario N2L 2Y5, Canada}
\setlength{\fboxsep}{0pt}

%\date{\today}
\begin{abstract}
The central puzzle of the cuprate superconductors at low hole density is the
nature of the pseudogap regime. It has a number of seemingly distinct experimental signatures: 
a suppression of the paramagnetic spin susceptibility at high temperatures, low energy electronic excitations that extend over
arcs in the Brillouin zone, X-ray detection of charge density wave order at intermediate temperatures, and quantum oscillations
at high magnetic fields and low temperatures. 
We show that a model of competing charge density wave and superconducting orders provides
a unified description of the intermediate and low temperature regimes. 
We treat quantum oscillations at high field beyond semiclassical approximations, and find clear and robust signatures of an electron pocket compatible with existing observations; we also predict oscillations
due to additional hole pockets. In the zero field and intermediate temperature regime, we compute the electronic spectrum in the presence of thermally fluctuating charge density and superconducting orders. Our results are compatible with experimental trends.
\end{abstract}
\maketitle

The suppression of the paramagnetic spin susceptibility of the cuprate superconductors at a high
temperature \cite{alloul} (often denoted $T^\ast$) implies the growth of antiferromagnetic spin correlations,
and a gap-like decrease in the electronic density of states at the Fermi level. It is useful to separate
existing theoretical models of the resulting pseudogap regime into two broad categories.

In the first category, the antiferromagnetic correlations signal the onset of a 
quantum spin liquid  \cite{YRZ,Wen12,Balents05,PS12,DCSS14b,Johnson11}, which in turn could be unstable to other symmetry-broken phases at lower temperatures.
For this to be a useful characterization, there should be remnants of the topological order of the spin liquid
at high temperatures. One possibility is the presence of closed Fermi pockets which violate the Luttinger theorem
constraining the total area enclosed by the Fermi surface \cite{YRZ,Wen12,PS12,DCSS14b}, 
but so far photoemission spectra only show intensity
on open arcs in the Brillouin zone.

In the second category \cite{KivelsonRMP,SSRMP,pines,kampf,so5}, the antiferromagnetic correlations are precursors to the appearance
of antiferromagnetism, superconductivity, charge density wave, and possibly 
other conventional orders at low temperatures. In the pseudogap regime, we then have primarily thermal and classical, rather than quantum,
fluctuations of these orders. 
In the recent work of Hayward \etal \cite{SSAF13}, it was shown that the unusual temperature dependence
of the X-ray scattering signal of the charge density wave order \cite{MHJ11,Ghi12,DGH12,SH12,MHJ13} is obtained naturally from an effective classical
model of angular thermal fluctuations of the charge-density wave and superconducting orders alone. The same model 
has also been connected
to diamagnetism measurements over the same temperature range \cite{SSAF14}. Within this framework,  in the intermediate temperature range over which the charge order correlations have been observed, antiferromagnetic correlations need not be included explicitly, but can be absorbed into the phenomenological parameters of the classical model of charge and superconducting orders. 

In the underdoped regime, quantum oscillations have been observed in two different families of the cuprates at high-fields and low temperatures \cite{LT07,leboeuf,louis2,harrison,suchitra2,greven,suchitra3}. The results seem to indicate that the phase at high-fields has Fermi-liquid like properties, albeit with a Fermi-surface that occupies only a tiny fraction of the Brillouin zone.
Harrison and Sebastian propose \cite{harrison} that these can be understood as a consequence of a bidirectional charge density
wave order: they argue that with the Fermi surface topology of the hole-doped cuprates, such an order leads to oscillations from a single
electron pocket, and that this is compatible with all important observed features. Their analysis is based upon a 
computation in zero field, followed by a semiclassical account of the influence of the magnetic field. However, since incommensurate charge order induces
a complex Fermi surface reconstruction, it is not a priori evident that this particular electron pocket would dominate the oscillations. 

In this work, we demonstrate that a model of charge-density wave and superconducting orders provides a unified
description of the high-field, low-temperature quantum-oscillations, and of certain features of the zero-field photoemission results at intermediate temperatures. While this model fits naturally into the second category of theories of the pseudogap summarized above, it may also be accommodated by the symmetry breaking instabilities of the spin liquid models in the first category.
First, we present a fully quantum mechanical analysis of the oscillations, carried out in a model which includes the lattice potential, charge order, and the magnetic field. We do indeed find signatures
of the electron pocket in the quantum oscillations. We also find clear oscillations from smaller hole pockets, which should
be detectable in experiments. Our analysis includes a description of the crossover in the oscillations from bidirectional (checkerboard) to unidirectional (stripe-like) density wave order. Second, we turn to photoemission experiments \cite{ZXS03,kanigel06, PJ08,Hashimoto12, Inna12, Kaminski13,Dessau12} at intermediate
temperatures. We couple electrons to the thermally
fluctuating charge and superconducting orders described by the angular fluctuation model of Hayward \etal \cite{SSAF13},
and compute the electron spectral function as a function of momentum and energy. Our results are in good agreement with some of the observed trends.
%%%%%%%%%%%%%%%%%%%%%%%%%%%%%%%%%%%%%%%%%%%%%%%%%%%%%
\begin{figure}[ht!]
\begin{center}
\includegraphics[width = \linewidth]{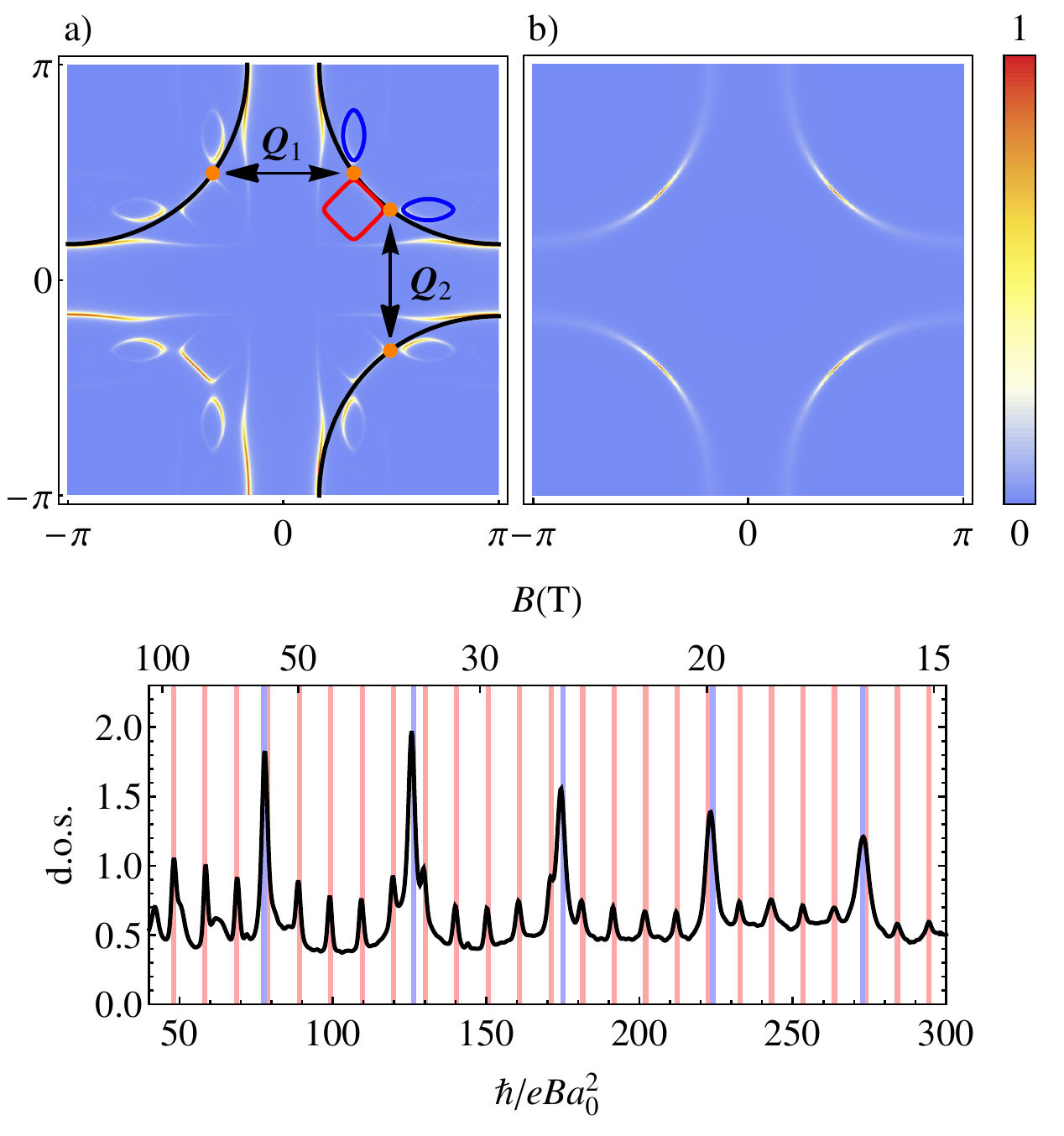}
\end{center}
%\rule{\linewidth}{1pt}
\caption{{\bf Spectral functions in the presence of static and fluctuating order.} (a) The color density plot displays the electron spectral function in the presence of long-range bidirectional bond density wave (BDW) at zero magnetic field (in the unfolded Brillouin zone). Such long-ranged BDW is likely to be present only in a strong magnetic field and will not be seen in ARPES experiments. Annotations are superimposed to highlight aspects of the spectral density. In black, the Fermi surface used for our computation. Dashed arrows mark the wavevectors of the BDW.  The BDW causes reconstruction of the Fermi surface and the formation of an electron-like pocket, marked in red, and two hole-like pockets, marked in blue. The pocket contours are obtained by semiclassical analysis as described in sec.~\ref{sec:semiclassical_approximation}.
The parameters are $t_1 = 1.0$, $t_2 = -0.33$, $t_3 = 0.03$, $\mu = -0.9604$, $p = 10\%$, $P^x_0 = P^y_0 = 0.15$, $\delta = 0.317$. (b) Electron spectral function in the presence of fluctuating superconducting and bond density wave correlations. The parameters are $p = 11\%$, $\Delta_0 = P_0^x = P_0^y = 1$, $T/t_1 = 0.06$, $g/\Lambda^2  = 0.2$, $\rho_\tn{S} = 0.05$, $\Lambda = 2$. The details are discussed in sec. \ref{sec:fluctuating_order}
(c) Quantum oscillations in the density of states induced by an applied magnetic field: red lines mark peaks associated with the electron pocket (frequency 432 T or 1.55\% of Brillouin zone), and blue lines those from the hole pockets  (frequency 90.9 T or 0.326\% of Brillouin zone).}
\label{fig:oscillations_summary_sf}
\end{figure}

\section{Results}
\subsection{Model} 
\label{sec:model}

We base our analysis on the following model hamiltonian
\begin{align}\label{eq:model_hamiltonian}
\begin{split}
H = &\sum_{\vec r, \vec a}\bigg[-t_a c_{\vec r + \vec a }^\dagger c^{\phantom{\dag}}_{\vec r} + \Delta_{\vec a} \Psi_{\vec r + \vec a /2} c_{\vec r + \vec a,\uparrow}^\dagger c_{\vec r,\downarrow}^\dagger  + \tn{h.c.}\\
&  + \sum_i P^{i}_{\vec a} e^{\ii \vec Q_i \cdot (\vec r + \vec a / 2) } \Phi^{i}_{\vec r + \vec a / 2} c_{\vec r + \vec a}^\dagger c^{\phantom{\dag}}_{\vec r} + \text{h.c.}\bigg]\,.
\end{split}
\end{align}

Here $\vec r$ labels the sites of a square lattice and the vector $\vec a$ runs over first, second and third neighbors, and also on-site $(\vec a = 0)$. The first term is the usual kinetic term, with hopping parameters $t_{a}$. 

The second term couples the electron to the superconducting order parameter. The coefficient $\Delta_{\vec{a}}$ specifies the superconducting form factor and the field $\Psi$ is the superconducting order parameter: it can be short-ranged or  acquire an expectation value.

The third term couples the fermion to the bond order. The index $i$ labels different wavevectors $\vec Q_i$, the coefficients $P^{i}_{\vec{a}}$ specify the corresponding form factors and the fields $\Phi^i$ are the order parameters, which can also be long-ranged or fluctuating. Throughout this work we consider a specific form of a density wave which resides primarily upon the bonds of the 
lattice: this is not crucial for the quantum oscillations, but is important for the electron
spectral function at intermediate temperatures. Building on recent experimental and theoretical work \cite{MMSS10,SSRP13,Efetov12,Metzner1,Metzner2,Yamase,Kee,Kampf,DHL13,JSSS14,AASS14a,AASS14b,KE13,comin2,SSJSD14} 
we use a bond density wave (BDW) with a $d$-form factor.

Both interaction terms can be obtained by appropriate decoupling of the Heisenberg interaction in the particle-hole and particle-particle channels \cite{SSRP13}.

We use a $d$-wave superconducting form factor $\Delta_{\pm \vec{\hat x}} = +\Delta_0 / 2, \Delta_{\pm \vec{\hat y}} = -\Delta_0 / 2$ and a bond order with the same form factor $P^{i}_{\pm \vec{\hat x}} = +P^i_0/ 2, P^{i}_{\pm \vec{\hat y}} = -P_0^i/ 2$ \cite{MMSS10,SSRP13} which is supported by recent experimental evidence \cite{comin2,SSJSD14}
although a small $s$-wave component may also be present \cite{DCSS14}. We consider a set of two wavevectors $\Q_1= 2 \pi (\delta,0)$ and $\Q_2=2 \pi (0,\delta)$, with $\delta \sim 0.3$, also based on experimental evidence.

A summary of our main results appears in fig.\ref{fig:oscillations_summary_sf}, which shows the electronic spectral functions in the presence of long-range incommensurate BDW (fig.\ref{fig:oscillations_summary_sf}a) and in the presence of fluctuating BDW and superconductivity (fig.\ref{fig:oscillations_summary_sf}b).

\subsection{Quantum oscillations from density wave order}
\label{sec:quantum_oscillations}

\begin{figure}
\includegraphics[width=\linewidth]{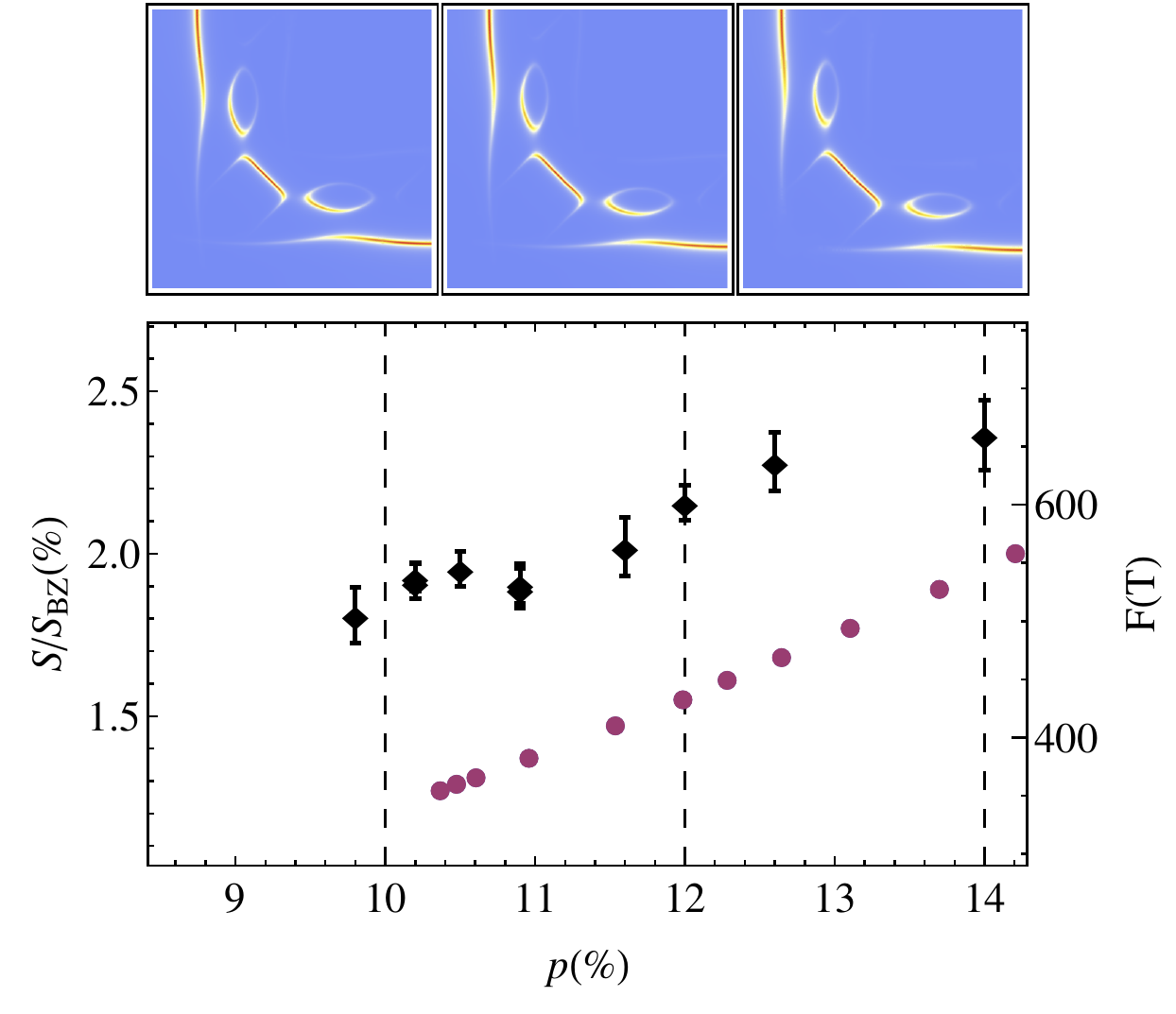}
\caption{\label{fig:doping_dependence} {\bf Doping dependence of the electron pocket oscillation frequency.} The computed frequencies of the electron pockets (red dots), in comparison with experimental data (black diamonds, adapted from \cite{vignolle}). Oscillations from the hole pocket have frequency $\sim$100T, and are off-scale. On the top, electron spectral function corresponding to $p = 10\%,\,12\%,\,14\%$. The commensuration factor $\delta = 0.40 - 0.73 p$, extrapolated from measurements in \cite{blackburn}; if we slightly modify this relation to $\delta = 0.39 - 0.73 p$, we obtain oscillation frequencies much closer to experimental observations. $P^x_0 = P^y_0 = 0.15$.
}
\end{figure}

\begin{figure}
\includegraphics[scale=0.5]{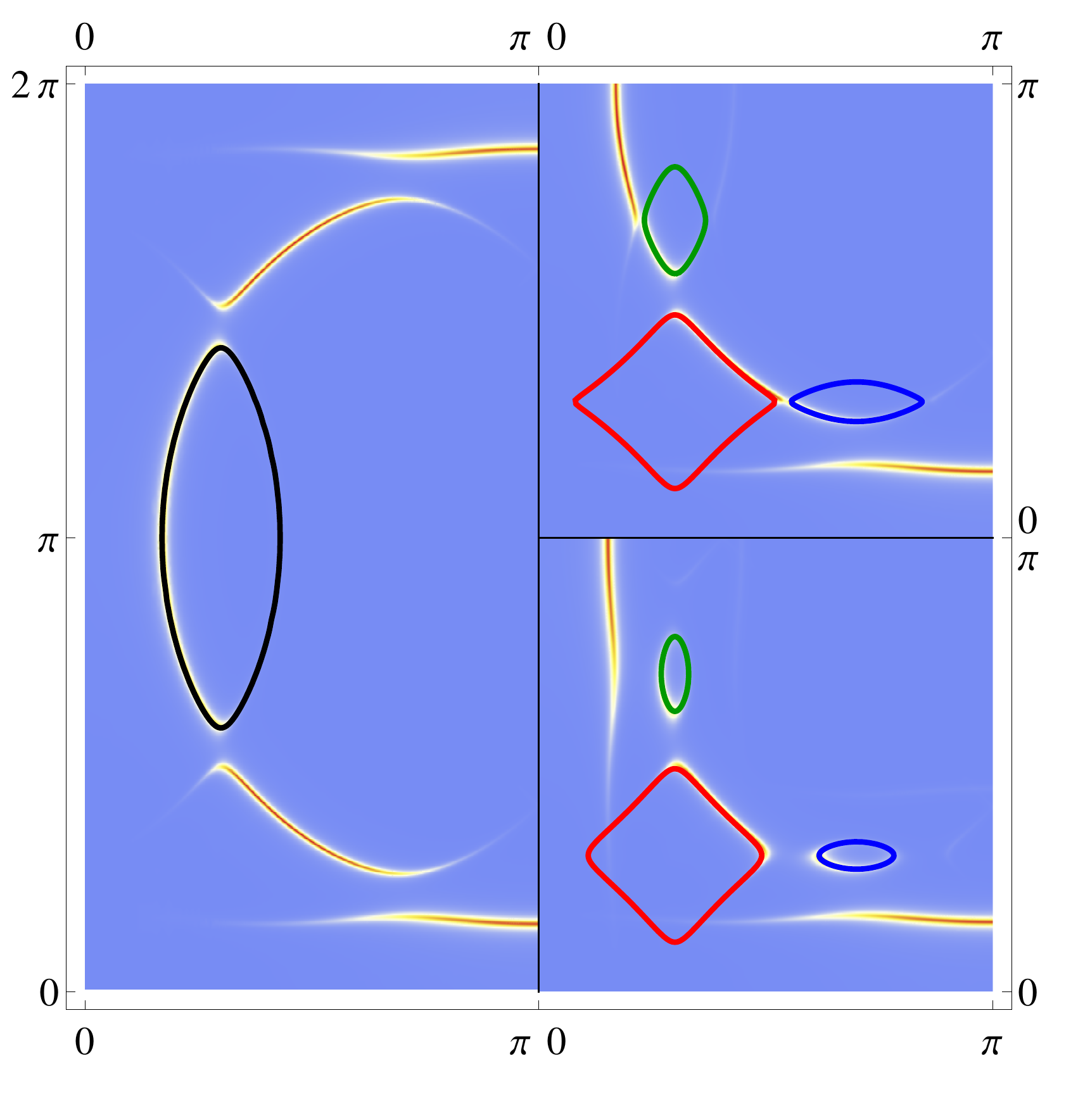}
%\rule{\linewidth}{1pt}
\caption{\label{fig:stripe_to_checkerboard_sf} {\bf Evolution of $A(\k,\omega=0)$ with increasingly isotropic order.} Electron spectral function with resonant orbits in evidence, as the order transitions from stripe to checkerboard: stripe order ($P^y_0 / P^x_0 = 0$) on the left, anisotropic checkerboard ($P^y_0 / P^x_0 = 0.2$) top right, isotropic checkerboard ($P^y_0 / P^x_0 = 1$) bottom right. $p = 11\%$, $P^x_0 = 0.15$, $\delta = 0.3$.}
\end{figure}

\begin{figure*}
\begin{center}
\includegraphics[width=\linewidth]{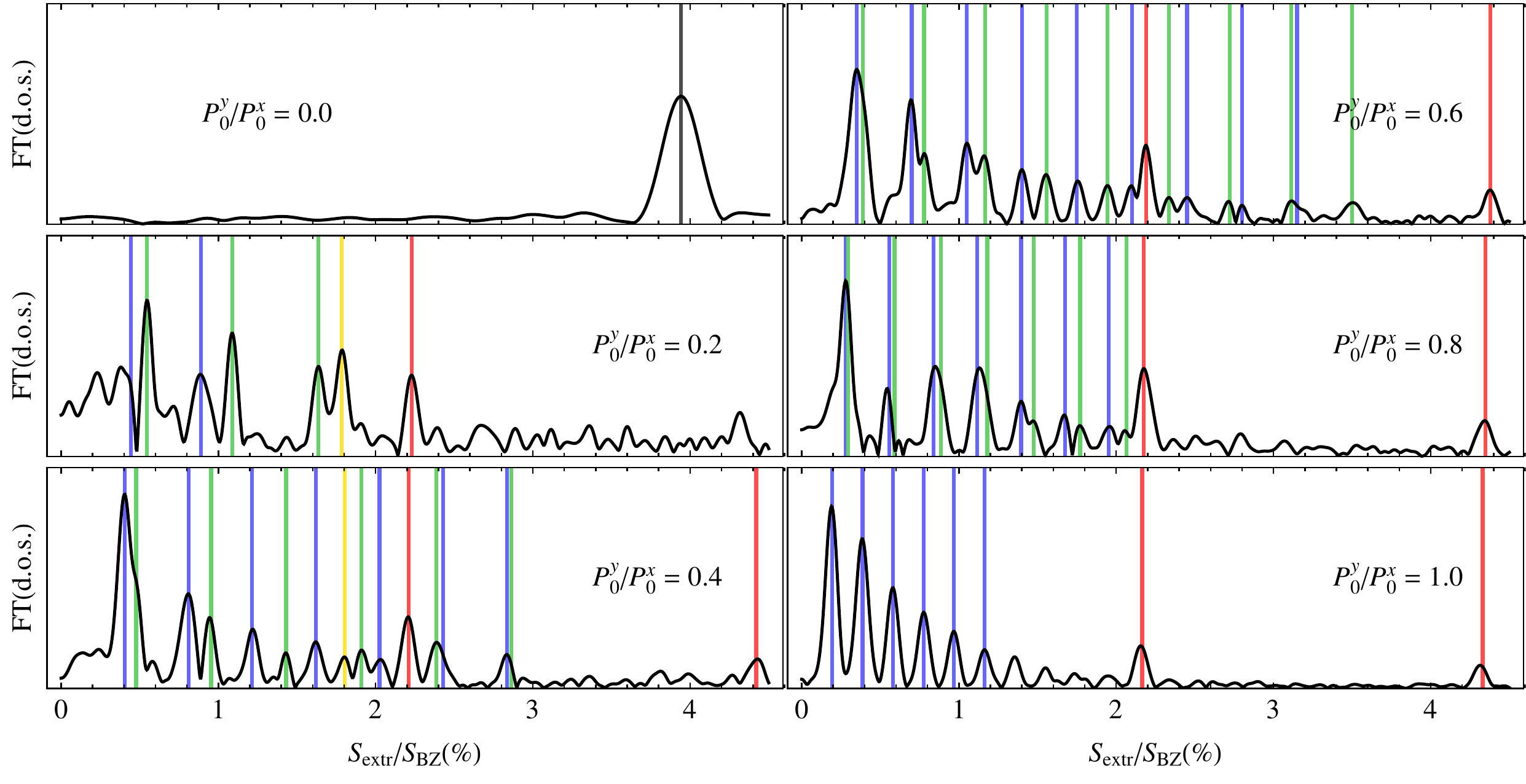}
\end{center}
\caption{\label{fig:stripe_to_checkerboard_osc} {\bf Fourier transform of the density of states with increasingly isotropic order.} 
$P_0^y /P_0^x = 0.0$ corresponds to the pure stripe state, while $P_0^y/P_0^x= 1.0$ is the state in which the density wave has equal
amplitude along the $x$ and $y$ directions. 
Colored vertical bands associate every peak with the corresponding orbit in Fig.~\ref{fig:stripe_to_checkerboard_sf} and~\ref{fig:oscillations_summary_sf}. Multiple bands with the same color denote higher harmonics of the same fundamental. The golden band is associated with the magnetic breakdown process shown in supplementary figure~1. $p = 11\%$, $P^x_0 = 0.15$, $\delta = 0.3$.}.
\end{figure*}

Fig.~\ref{fig:oscillations_summary_sf}(a) illustrates Fermi surface reconstruction by a bi-directional density wave modulation with wavevectors parallel to the Cu-O bonds and period $\sim 3$ lattice spacings. We constrain the Fermi surface by spectroscopy experiments \cite{Hossain2008}, and use the remaining freedom in hopping parameters to obtain an electron pocket of about the right size. While our analysis mainly centers around the experiments on YBCO, which has a sligthly orthorombic lattice, for simplicity we assume tetragonal symmetry. We also ignore the effect of bilayer splitting. Based on the results of Sebastian \etal \cite{suchitra4}, we expect that the inclusion of these effects would produce additional satellite frequencies, but we leave these issues to future work.

The BDW opens up gaps at nested points on the Fermi surface which gets reconstructed into a pocket, highlighted in red, of about the right size, carrying electron-like (negatively charged) excitations. This reconstruction scheme was first proposed in \cite{harrison} as a way to explain the small quantum oscillation frequency observed in underdoped YBCO \cite{LT07}, as well as negative Hall and Seeback coefficients \cite{leboeuf} observed at low temperature in high magnetic fields.  Quite generically, this reconstruction scheme also produces two hole pockets, marked in blue in the figure. 

In the background of Fig.~\ref{fig:oscillations_summary_sf}(a) we display the electron spectral function. In this observable, the four sides of the electron pocket are manifest as four disconnected arc-like features near the nodal regions. In a semiclassical picture, these four arcs are recomposed in a single orbit as described in section~\ref{sec:semiclassical_approximation}.
Note that the spectral function is obtained in the presence of long-range charge order. At zero field the order is short-ranged and the effect is less visible. Therefore, it would be quite difficult to observe signatures of the electron pocket in photoemission experiments.

The presence of closed pockets causes oscillations in the density of states at the Fermi level, shown in Fig.~\ref{fig:oscillations_summary_sf}(c).
Periodic oscillations with frequency matching the area of both the electron and the hole pockets are clearly recognizable,  corresponding to two pockets of area 1.55\% and 0.326\% of the Brillouin zone. We find matching semiclassical orbits as follows: the red and blue contours in Fig.~\ref{fig:oscillations_summary_sf}(a) are the contours $E_{\vec k} = 0$ discussed in \ref{sec:semiclassical_approximation}. By numerical integration, we find the area within the red contour to be 1.45\% of the Brillouin zone, and we match it to the shortest oscillation period, while the area within the blue contour is 0.321\% of the Brillouin zone, corresponding to the largest period. Both semiclassical frequencies are quite close to the exact result.

We also note that strictly speaking, when the order is incommensurate, the number of bands crossing the Fermi energy is infinite. Even in the presence of commensurate order, but with a long period, the tangle of bands is quite dense. For this reason, we find it more convenient to plot the spectral function in the full Brillouin zone as it clearly shows the bands that carry the highest spectral weight. Moreover, the features highlighted by the spectral function are reflected quantitatively in the quantum oscillations observed in our computations. 
As the strength of the incommensurate order is increased, it has been argued \cite{YZAMSK14} that the incommensurate nature will play a more significant role in disrupting the oscillations.

Let us stress that previous treatments of the oscillations in the cuprates were restricted to charge modulations with commensuration periods of 3-4 lattice spacing, whereas the experimental BDW is incommensurate. The BDW discussed  in Fig.~\ref{fig:oscillations_summary_sf} has a commensuration period of 10 lattice spacings and $\delta = 0.3$, close to experimental values, and there is no obstruction to studying even longer commensuration periods. In fact, we are able to compute the doping dependence of the oscillation frequency as the commensuration factor $\delta$ varies. Although there are theoretical models \cite{MMSS10, SSRP13} predicting that the bond order wavevector is tied to the geometry of the Fermi surface, connecting special points called hot spots where the Fermi surface intersects the magnetic Brillouin zone, it is unclear if the experimentally obtained wavevector is identical to this wavevector across different families of the cuprates. In this work, we prefer to obtain the wavevector from experiment, and we take $\delta = 0.40 - 0.73 p$, based on the data in ref.~\cite{blackburn}. 

As shown in fig.~\ref{fig:doping_dependence}, the frequency of oscillations associated with the electron pocket grows with doping, in reasonable agreement with experiments. This is confirmed by the growing size of the pocket as seen in the spectral function. We also find that the size of the hole pocket decreases with doping, and we observe a corresponding change in the oscillation frequency.While this very small pocket is a rather natural and almost unavoidable consequence of this reconstruction scheme, the detection of corresponding quantum oscillations is challenging. In particular, it may be difficult to distinguish them from the interference between electron pocket oscillations due to bilayer splitting. Moreover, its location closer to the antinodal region leaves it more exposed to enhanced scattering and superconducting fluctuations, and this could suppress the amplitude of oscillations beyond detection.  There are however indications of its existence in the recent work of Ref.~\onlinecite{cyril}.

The reconstruction scheme used so far assumes a bidirectional bond density wave with equal amplitude on both wavevectors $\vec Q_1 = 2\pi\, (\delta,\, 0)$ and $\vec Q_2 = 2\pi\, (0,\, \delta)$, a so called checkerboard order. As there is some experimental evidence \cite{nematic1,nematic2} that instead supports stripe order, it is natural to ask how the picture changes when the order is not isotropic. Other analysis of quantum oscillations in the presence of stripe order are in Refs.~\onlinecite{MN07,YLK11,EWC12}, but with models and Fermi surface configurations which differ from ours.

Fig.~\ref{fig:stripe_to_checkerboard_sf} displays the electron spectral function for purely stripe order, anisotropic checkerboard order, and isotropic checkerboard order. When the order is unidirectional, a single large ($\sim 4\%$ of Brillouin zone, in black) hole pocket is generated. Turning on a small amplitude on the second wavevector causes the opening of further gaps, and the creation of one electron pocket (in red) and two inequivalent hole pockets (blue and green), which are smoothly connected to the ones present in the isotropic case. All these features have a consequence on the spectrum of quantum oscillations, as shown in Fig.~\ref{fig:stripe_to_checkerboard_osc}. In particular, oscillations from the large hole pocket are disrupted as soon as the second wavevector gains a relatively small amplitude, and peaks associated with the three smaller pockets are immediately visible.

There is only one major peak whose frequency does not correspond to the area of any pocket, and it is marked by a golden band. This peak is a manifestation of magnetic breakdown, and is discussed in more detail in the supplementary notes 1.

\subsection{Order parameter fluctuations above $T_\tn{c}$}
\label{sec:fluctuating_order}

A quantitative effective description of the zero field, $T > T_\tn{c}$ regime of the underdoped cuprates has recently been proposed in terms of a multicomponent O(6) order parameter, $\vec n$ (a 6-component vector of unit length), collecting the order parameters of both superconductivity and BDW \cite{SSAF13, MMSS10, Efetov12,Fradkin10}. 
At low temperature, terms that break the O(6) symmetry explicitly cause $\vec{n}$ to fluctuate preferentially along the direction that corresponds to superconductivity. 
This produces long-range superconductivity.  On the other hand, at higher temperature, O(6) symmetry is approximately restored and $\vec{n}$ fluctuates along all directions, yielding short-ranged superconducting and bond order correlations. 
Here, we couple the O(6) order to the electrons and compute the electronic self energies. Similar computations of photoemission spectrum have been carried out earlier for the case of superconducting fluctuations \cite{EBEA07,TSPAL09,SBTVR11,SBMR13,norman}; our analysis below shows that a combined model agrees well with many of the observed trends as a function of temperature, angle around
the Fermi surface, and energy.

The O(6) field, $\vec n$, collects all components of the superconducting and bond order parameters,
\begin{equation}
 \vec n = (\Re \Psi,\, \Im \Psi, \,\Re \Phi^x,\, \Im \Phi^x,\, \Re \Phi^y,\, \Im \Phi^y)
\end{equation} 
and is governed by a non linear $\sigma-$model (NL$\sigma$M) action:
\begin{align}
&Z=\int{\cal{D}}\vec{n}(\vec r)\delta(\vec{n}^2-1)~\tn{exp}\bigg(-\frac{\cal{S}_\tn{cl}}{T}\bigg),\\
\begin{split}\label{act}
&{\cal{S}}_\tn{cl}=\frac{\rho_{\tn{S}}}{2}\int d^2r\bigg[|\nabla\Psi|^2+\lambda(|\nabla\Phi_x|^2 + |\nabla\Phi_y|^2)\\
&\phantom{{\cal{S}}_\tn{cl}=}+g(|\Phi_x|^2 + |\Phi_y|^2)+w(|\Phi_x|^4 + |\Phi_y|^4)\bigg]\,,
\end{split}
\end{align}
where $\rho_{\tn{S}}$ and $\rho_{\tn{S}}\lambda$ control the helicity moduli of the superconducting and of the density wave order respectively. The coupling $g$ breaks the symmetry between the $\Psi$ and $\Phi_x, \Phi_y$ directions. It sets the relative energetic cost of ordering between the superconducting and density wave directions. We take $g > 0$, so that superconductivity is preferred at low temperatures (see supplementary figure 2). The coupling $w$ imposes the square lattice point group symmetry on the density wave order.

In the large $N$ limit ($N$ being the number of components of $\vec n$), the propagators of $\Psi$ and $\Phi_{x,y}$ have a massive form of the type (we ignore here the possibility of having a $C_4\rightarrow C_2$, Ising symmetry breaking),
\begin{align}
 &D^\tn{s}_{\tn{cl}}(\vec q)=\frac{1}{\vec q^2+\overline\sigma},&& ~D^\tn{b}_{\tn{cl}}(\vec q)=\frac{1}{\lambda \vec q^2 + \overline\sigma+g+\overline\phi},
\end{align} 
where the parameters $\bar \sigma$ and $\bar \phi$ are to be determined in terms of the couplings in (\ref{act}) by solving the large $N$ saddle point equations,
\begin{equation}
\begin{split}
&\frac{\rho_{\tn{S}}}{T}=\int \frac{\dd^2 \vec q}{(2\pi)^2}\left[\frac{1}{\vec q^2+\overline\sigma} + \frac{2}{\lambda\vec q^2+\overline\sigma+g+\ol\phi} \right],\\
&\ol\phi=\frac{2wT}{\rho_{\tn{S}}}\int \frac{\dd^2 \vec q}{(2\pi)^2}\frac{1}{\lambda\vec q^2+\overline\sigma+g+\ol\phi}\,,
\label{splat}
\end{split} 
\end{equation} 
which we regulate with a hard momentum cutoff $\vec q^2 < \Lambda^2$ (see supplementary methods A).
%%%%%%%%%%%%%%%%%%%%%%%%%%%%%%%%%%%%%%%%%%%%%%%%%%%%%%%%%%%
\begin{figure*}
\begin{center}
\includegraphics[width=\linewidth]{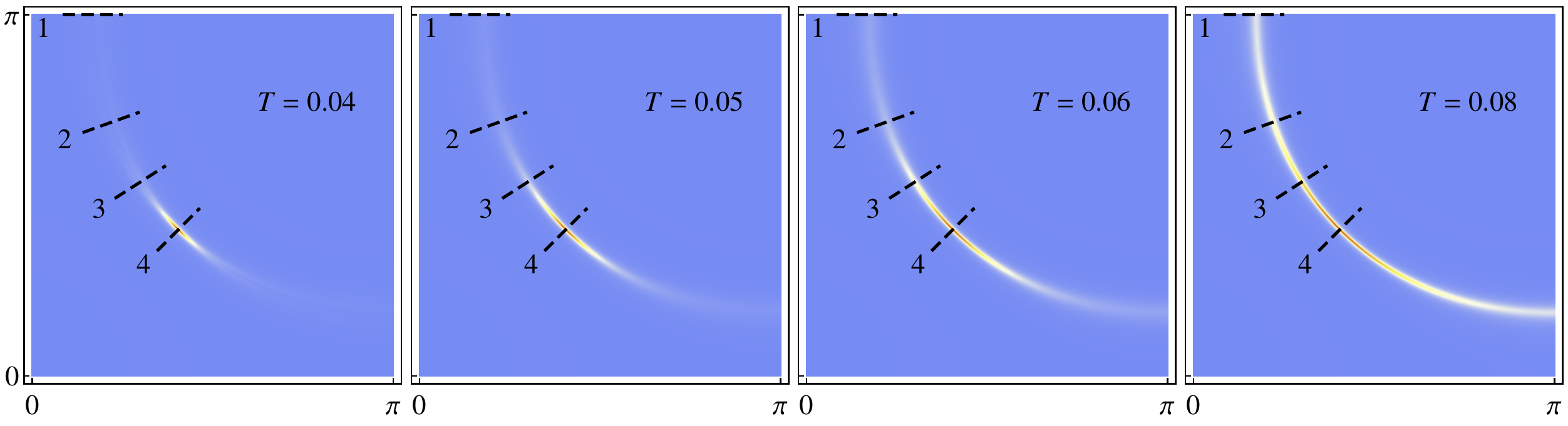}
% \rule{\linewidth}{1pt}
\end{center}
\caption{{\bf $A(\k,\omega=0)$ as a function of increasing temperature.} Electron spectral function in the presence of superconducting and bond order fluctuations, at four different temperatures. Dashed lines mark the cuts shown in Fig.~\ref{fig:cuts_across}. At sufficiently low temperatures superconducting fluctuations gap out the antinodal region. Bond order fluctuations are very short-ranged and do not have observable effects. $p = 11\%$, $\Delta_0 = P_0^x = P_0^y = 1$, $\rho_{\tn{S}} = 0.05$, $\Lambda = 2$.}
\label{akwevolution}
\end{figure*}
%%%%%%%%%%%%%%%%%%%%%%%%%%%%%%%%%%%%%%%%%%%%%%%%%%%%%%%%%%%

%%%%%%%%%%%%%%%%%%%%%%%%%%%%%%%%%%%%%%%%%%%%%%%%%%%%%%%%%%%
\begin{figure}
\begin{center}
\includegraphics[width=\linewidth]{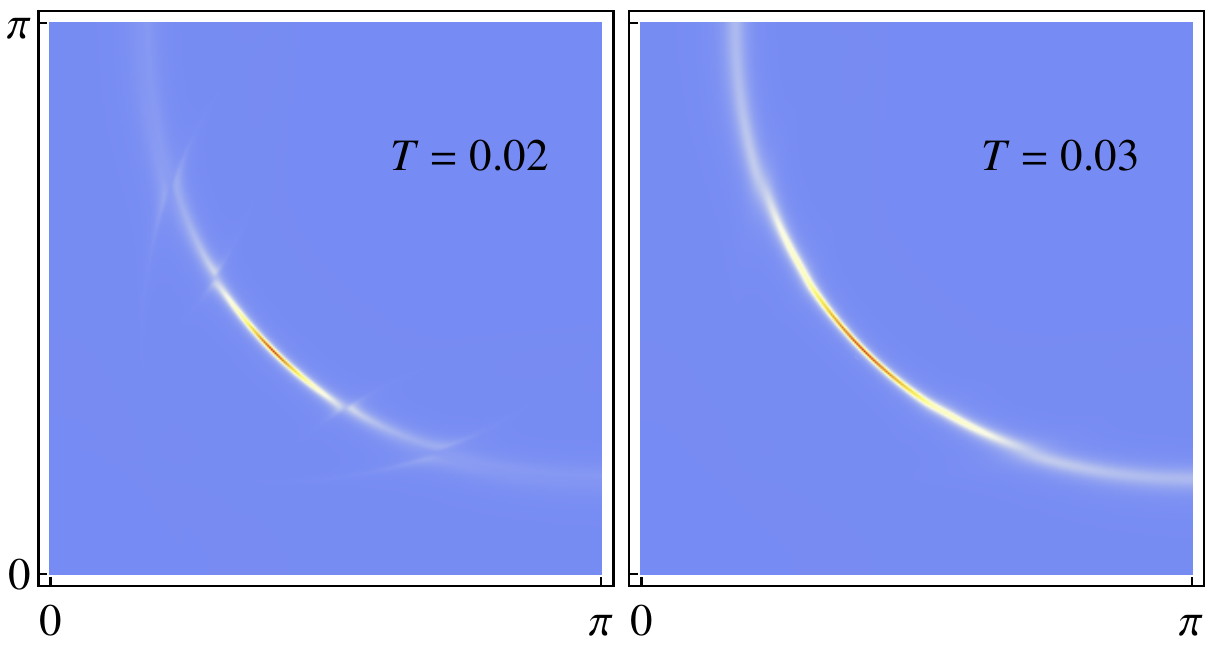}%
\end{center}
\caption{\label{fig:constant_frequency}{\bf $A(\k,\omega=0)$ for the fully isotropic O(6) model.} Electron spectral function in the presence of superconducting and bond order fluctuations for a fully isotropic O(6) model ($g/\Lambda^2 = 0$). At sufficiently low temperatures bond order fluctuations lead to incipient Fermi surface reconstruction. $p = 11\%$, $\Delta_0 = P_0^x = P_0^y = 1$, $\rho_{\tn{S}} = 0.05$, $\Lambda = 2$.}
\label{CDWshad}
\end{figure}
%%%%%%%%%%%%%%%%%%%%%%%%%%%%%%%%%%%%%%%%%%%%%%%%%%%%%%%%%%%

The mass $\ol\sigma$ of the superconducting order parameter  is strongly temperature dependent, and it is exponentially suppressed if $T\ll \rho_{\tn{S}}$:
\begin{equation}
  \frac{\ol\sigma(T)}{\Lambda^2}\sim e^{-4\pi\rho_{\tn{S}}/T}\,.
\end{equation} 
This is how the large $N$ sigma model approximates long-range superconductivity (see supplementary figure 3). On the other hand, for $T \gg \rho_{\tn{S}}$
\begin{equation}
 \frac{\ol\sigma(T)}{\Lambda^2} \sim \frac{3}{4\pi} \frac{T}{\rho_{\tn{S}}} - \frac{2}{3} \frac{g}{\Lambda^2} - \frac{1}{2} + \mathcal O\left(\frac{\rho_{\tn{S}}}{T}\right)\,.
\end{equation} 

The mass of the bond order parameter is $\bar \sigma + g$ (when $w=0$), and hence the parameter $g$ determines the correlation length of the bond order at low temperatures. We take the universal number $g/\Lambda^2 \sim 0.2$, based on the results in ref. \cite{SSAF13}, but we still have to fix $\Lambda$. We choose it in such a way that the correlation length of the density wave order at low temperature is a few lattice spacings. We also verify explicitly that our conclusions are not affected as the correlation length of the bond order varies from 2 to 10 lattice spacings, which is the range in which the experimental value lies.

The NL$\sigma$M (\ref{act}) is a completely classical model, and neglects the fact that superconducting and bond order parameters are coupled to the gapless fermionic degrees of freedom along the Fermi surface. We include this contribution at RPA level, computing the one-loop self-energy of the bosons due to the interaction with the fermions (see supplementary figure 4). The sole effect of these bubbles is to give rise to damping terms at low energies (see supplementary methods B). The improved form of the propagators is,
\begin{equation}
\begin{split}
&D_\tn{s}(\vec q,\,\ii \Omega_n)=\frac{1}{\alpha_\tn{s}|\Omega_n|+\vec q^2 + \bar \sigma}\,,\\
&D_\tn{b}(\vec q,\,\ii \Omega_n)=\frac{1}{\alpha_\tn{b}|\Omega_n|+\lambda\vec q^2 + \bar \sigma + g + \bar \phi}\,,
\end{split}
\end{equation} 
where $\alpha_\tn{s,b}$ parametrizes the strength of damping (we will take $\alpha_\tn{s} = \alpha_\tn{b} = 1$ in the rest of the computation).

The electron spectral function due to the retarded self-energy $\Sigma(\vec k, \omega)$ of the electrons in the presence of fluctuating superconducting and bond order parameters is given by,
\begin{equation}
 A(\vec k, \omega) = - \frac{1}{\pi} \Im \frac{1}{\omega - \epsilon_{\vec k} - \Sigma(\vec k, \omega)}\,.
\end{equation}

%%%%%%%%%%%%%%%%%%%%%%%%%%%%%%%%%%%%%%%%%%%%%%%%%%%%%%%%%%%
\begin{figure*}
\begin{center}
\includegraphics[width=\linewidth]{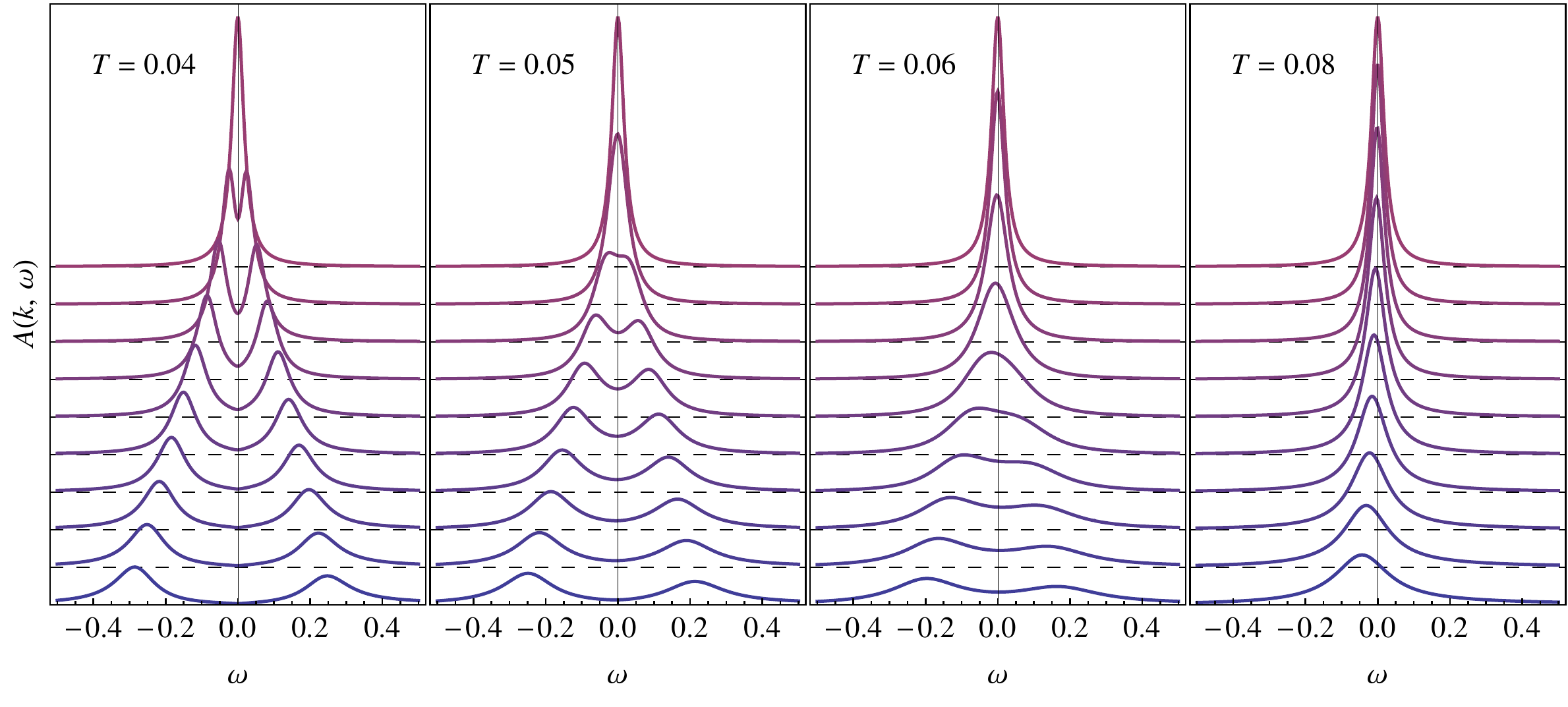}%
\end{center}
\caption{{\bf Electron spectral function as a function of frequency.} $A(\k,\omega)$ at several $\vec k$-points on the bare Fermi surface $\epsilon_{\vec k} = 0$: from the node in red at the top to the antinode in blue at the bottom. $p = 11\%$, $\Delta_0 = P_0^x = P_0^y = 1$, $\rho_{\tn{S}} = 0.05$, $g/\Lambda^2 = 0.2$, $\Lambda = 2$.}
\label{cutsakw}
\end{figure*}
%%%%%%%%%%%%%%%%%%%%%%%%%%%%%%%%%%%%%%%%%%%%%%%%%%%%%%%%%%%

The self-energy has two contributions, from the particle-particle and particle-hole channel. We have %, as shown in Fig.~\ref{self}. 
%\begin{widetext}
\begin{equation}
\begin{split}
&\Sigma_\tn{s}(\vec k,i\omega_m)= - \frac{1}{\beta}\sum_{n}\int_{\vec q}\ \D^2_{\vec k-\frac{\vec q}{2}} \frac{1}{-\ii \omega_m + \ii \Omega_n - \epsilon_{\vec q-\vec k}} \frac{1}{|\Omega_n| + \varepsilon^\tn{s}_{\vec q}}\,,\\
&\Sigma_\tn{b}(\vec k,i\omega_m) =+\frac{1}{\beta}\sum_{n, i}\int_{\vec q}\ P^{i\ 2}_{\vec k+\frac{\vec q}{2}}~\frac{1}{+\ii \omega_m + \ii \Omega_n - \epsilon_{\vec q+\vec k}}\frac{1}{|\Omega_n| + \varepsilon^\tn{b}_{\vec q - \vec Q_i}}\,,\\ 
\end{split}
\end{equation} 
%\end{widetext}
where $\Omega_n$ are bosonic Matsubara frequencies, $\omega_n$ are fermionic Matsubara frequenices and
\begin{align}
\begin{split}
 &\varepsilon^\tn{s}_{\vec q} = \vec q^2 + \bar \sigma\,,\\
 &\varepsilon^\tn{b}_{\vec q} = \lambda\vec q^2 + \bar \sigma + g + \bar \phi\,.
\end{split}
\end{align} 
The self energies are identical for up and down spin.

We carry out the Matsubara sum analytically (see the Methods section) and carry out the integral over $\vec q$ numerically using an adaptive integration routine.

There are two energy scales in the system: the bandwidth, set by the hopping $t_1\approx 3000$ K, and the bare helicity modulus $\rho_{\tn{S}}\approx 150 \rm{K}$, which controls the temperature at which there is an onset of phase fluctuations of $\vec n$ in the O(6) model, and hence the temperature dependence of the correlation length of superconducting and bond order fluctuations. In the following, we explore a range of temperature $\rho_{\tn{S}} \lesssim T \lesssim 2 \rho_{\tn{S}}$, with $\rho_{\tn{S}} = 0.05\, t_1$. We express temperature in units of $t_1$.

In Fig.~\ref{akwevolution} we display a section of the spectral function $A(\vec k,\omega)$ at constant frequency equal to the Fermi energy ($\omega = 0$), at four different temperatures. Even at the lowest temperature, a portion of the Fermi surface close to the node, a Fermi arc, survives as a contour of well-defined excitations in the presence of bond order and superconducting fluctuations. As the temperature increases, there is a gradual build up of spectral weight in the antinodal regions, and around $T\approx0.08$, the full Fermi surface is recovered. 

Let us now give an intuitive picture for what is causing the arcs. The damping of the superconducting order, combined with the $d-$wave form factor, leads to enhanced scattering of the fermionic excitations close to the anti-nodes. What is left in the form of an arc are essentially the nodal quasiparticles which have survived as sharp excitations. As temperature increases, a larger region around the nodes survives the effect of scattering by superconducting fluctuations. This is consistent with the observations of Kanigel et al. \cite{kanigel06}, who also observed an increase in the arc length as a function of increasing temperature. 

It is reasonable to expect that $d-$wave bond order fluctuations would also cause enhanced scattering around points of the Fermi surface that are nested by the 
BDW wavevector. However, the correlation length of the bond order is much shorter than the superconducting one, by merit of our choice of the parameters $g$ and $\Lambda$, and hence this phenomenon is not visible in Fig.~\ref{akwevolution}. In general, for correlation lengths of order a few lattice spacings, as seen in experiment, the effect of bond order fluctuations on the spectral function is negligible. 

If, however, the correlation length is enhanced, for example by setting $g = 0$, such that we have a fully isotropic O(6) model, then bond order fluctuations have a distinct effect, as shown in Fig.~\ref{CDWshad}. At the lowest temperature the original Fermi surface breaks up into an arc-like feature close to the nodes and a number of shadow-like features similar to those discussed earlier in connection to long-range bond order. In spite of the bond order correlations being short-ranged and fluctuating, the shadows arise purely from the real part of the self-energy. 

In addition to the observations made above, here we have the additional effect that superconducting fluctuations somewhat suppress the shadows near the antinodes. It would be interesting to look for signatures of these features in future experiments in high quality samples. 

%%%%%%%%%%%%%%%%%%%%%%%%%%%%%%%%%%%%%%%%%%%%%%%%%%%%%%%%%%%
\begin{figure*}
\begin{center}
\includegraphics[width=\linewidth]{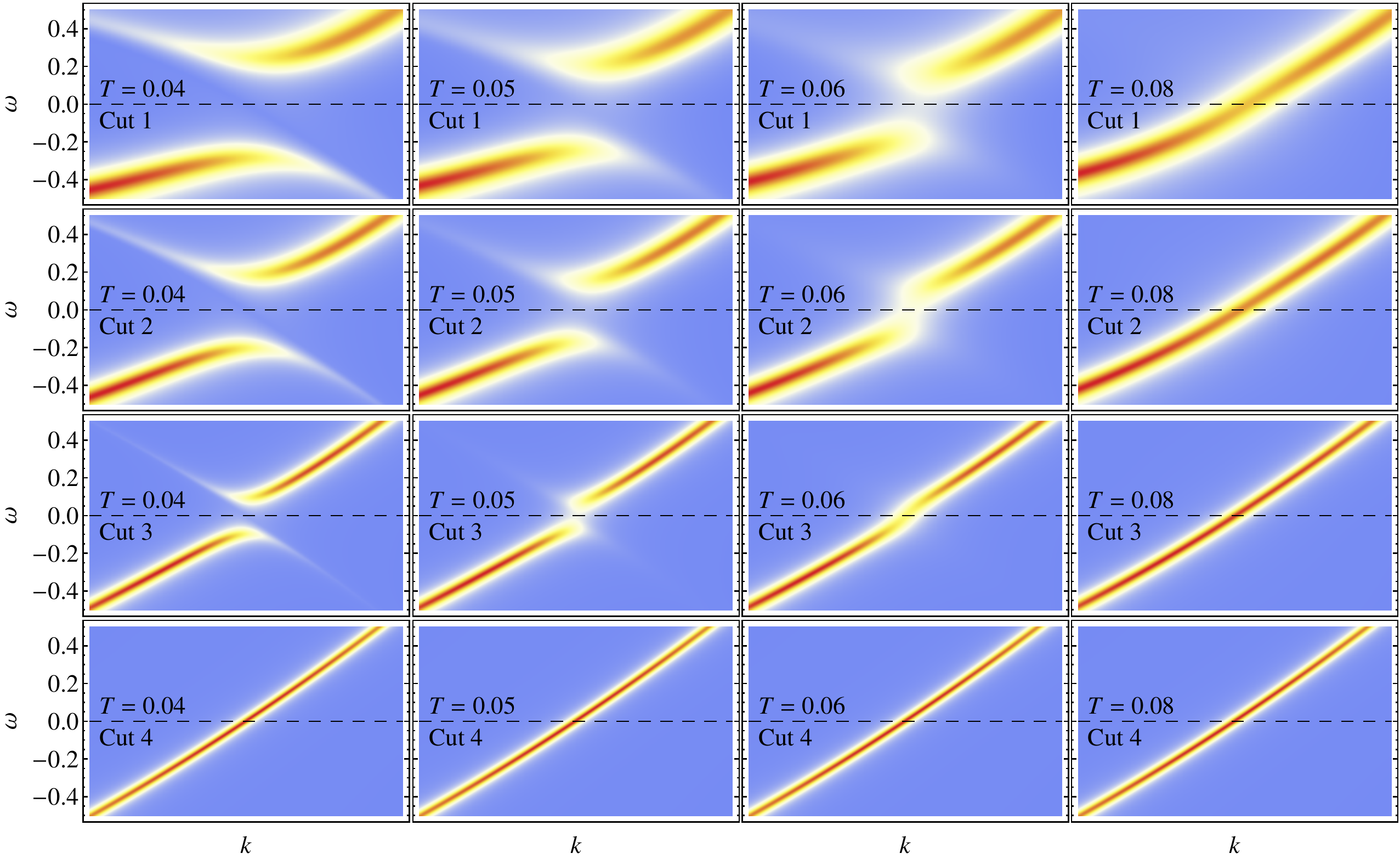}
\end{center}
\caption{\label{fig:cuts_across} {\bf Energy-momentum scans of $A(\k,\omega)$.} Frequency scans of the electron spectral function along the dashed lines in Fig.~\ref{fig:constant_frequency}, at four different temperatures. $p = 11\%$, $\Delta_0 = P_0^x = P_0^y = 1$, $\rho_{\tn{S}} = 0.05$, $g/\Lambda^2 = 0.2$, $\Lambda = 2$.}
\label{cutsw}
\end{figure*}
%%%%%%%%%%%%%%%%%%%%%%%%%%%%%%%%%%%%%%%%%%%%%%%%%%%%%%%%%%%
Interestingly, more recent experiments carried out using a different protocol seem to suggest that the arc length stays constant over a wide range of temperature and eventually undergoes a collapse at low temperatures \cite{Kaminski13}. This was presented as evidence for the BDW fluctuations playing a more prominent role in giving rise to the arcs. However, within our setup, the density wave
fluctuations with a correlation length of even up to 10 lattice spacings don't play any significant role in the arc phenomenology. Furthermore, in the presence of purely 
density wave fluctuations, the energy gap is centered above the Fermi energy for points situated between the antinodes and regions that are nested by the 
density wave wavevector \cite{Chubukov14}. This has been confirmed in ARPES experiments \cite{PJ08,Inna12}.

Let us now discuss the properties of $A(\vec k, \omega)$ as a function of frequency for $\vec k$ points on the bare Fermi surface $\epsilon_{\vec k} = 0$. As shown in Fig.~\ref{cutsakw}, at the lowest temperature $T=0.04$, a well-defined gap is present at the antinodes. Moving towards the node (from the blue to the red scans), the gap closes at a $\vec k$ point which we identify with the tip of the arc. As temperature increases the gap closes and the full Fermi surface is recovered at sufficiently high temperatures. We note in passing that the location of the peak at the antinode at $T=0.08$ is shifted away from $\omega=0$ due to a renormalization of the bare dispersion by $\Re \Sigma(\vec k,\omega=0)$.  

Fig.~\ref{cutsw} displays the quasiparticle dispersion along the black dashed lines labelled 1-4 in Fig.\ref{akwevolution} as a function of temperature. For the first cut, at the antinode, there is a well-defined gap in the quasiparticle spectrum at the Fermi surface, which closes as temperature increases. The overall scale of the gap is set by the magnitudes of $\D_0$ and $P_0^i$. A similar behavior is seen for the second and third cut, albeit with a smaller gap that closes at a lower temperature. The fourth cut is a scan across the nodes and the dispersion remains gapless since neither the SC nor the BDW fluctuations affect the nodal quasiparticles.

As noted above, for our specific choice of parameters, the BDW fluctuations are subdominant to the SC fluctuations in the photoemission spectrum. In particular, the dispersions and the gap structure in the normal state obtained from this analysis are approximately particle-hole symmetric (as evident from 
Figs.~\ref{cutsakw} and~\ref{cutsw}). However, recent experiments have seen evidence for particle-hole asymmetry at the tips of the Fermi-arcs and in the antinodes \cite{PJ08, Hashimoto12}. Such an asymmetry arises naturally in the presence of more pronounced BDW fluctuations (as compared to the superconducting fluctuations). Our formalism can reproduce some of these features in the presence of fluctuating BDW with longer correlation lengths and stronger coupling to the fermions.

Finally, we reiterate the statement in the introduction that the present computation is expected to apply at the 
intermediate temperatures where charge order fluctuations are appreciable. A more complex model also including antiferromagnetism
is likely needed at higher temperatures, which we do not discuss here.

\section{Discussion}
\label{conc}

We have analyzed the influence of incommensurate density wave order on quantum oscillations and the combined effect of pairing and density wave fluctuations on photoemission in the underdoped cuprates. In combination with other recent results \cite{SSAF13}, 
such a model appears to provide a satisfactory description
of the underdoped cuprates in current high-field \cite{LT07,leboeuf,louis2,harrison,suchitra2,greven,suchitra3}, photoemission \cite{kanigel06,Inna12,Kaminski13}, NMR \cite{MHJ11,MHJ13}, 
STM \cite{SSJSD14,nematic1,nematic2,JH02,aharon,AY04,comin13,neto13}, and X-ray \cite{Ghi12,DGH12,SH12,comin2,comin13,neto13} experiments. It can also be subjected to further tests by experiments at higher fields, in cleaner samples, or with
higher precision. However, our analysis does not strongly constrain the nature of the high temperature pseudogap, where we need sharper
experimental tests of the distinction between the two categories of models described in the introduction.
On this issue, a recent work \cite{DCSS14b} has argued that the wavevector of the $d$-form factor BDW supports
the spin liquid models in the first category. We suggest that unravelling the nature of the quantum-critical point apparent in recent
experiments near optimal doping \cite{JH14,JSD14,Ramshaw14}, and deciphering the nature of the strange metal, may be the route to resolving these
long-standing and difficult questions.

\section{Methods}
\subsection{Semiclassical treatment of quantum oscillations}
\label{sec:semiclassical_approximation}
Quantum oscillations are a manifestation of Landau quantization. As the magnetic field varies, Landau levels are brought across the Fermi energy one after the other, causing the density of states to peak periodically, with consequent oscillations in thermodynamic and transport quantities. We now briefly derive Landau quantization for free electrons on a lattice at the semiclassical level. The derivation is then extended to include the presence of periodic charge modulations, which may or may not be commensurate. The fully quantum computation will be carried out in the next subsection. For simplicity we work in 2 dimensions.

We introduce a wavepacket
\begin{equation}\label{eq:wavepacket}
 \ket{\psi(t)} = \sum_{\vec r} \exp\left[\ii \vec{k}(t) \cdot \vec r -\frac{(\vec{\bar r}(t) - \vec r)^2}{2 L^2}\right] \ket{\vec r}\,.
\end{equation} 

In order to derive an effective action for $\vec k$, $\vec{\bar r}$ we start from the action
\begin{equation}\label{eq:lattice_action}
 S = \int \dd t\, \bra{\psi} \left(\ii \partial_t - h \right) \ket{\psi}\,,
\end{equation} 
where
\begin{align}
 &h = -\sum_{\vec r, \vec a} t_{\vec a} \exp\left[\ii \int_{\vec r}^{\vec r + \vec a} \vec A\cdot \dd \vec l \right] \ket{\vec r + \vec a}\bra{\vec r}
\end{align}
is the single particle hamiltonian. Were $\psi$ the most general wavefunction, this action would yield the Schr\"odinger equation. Taking instead the parametrization (\ref{eq:wavepacket}), substituting and neglecting subleading terms in $1 / L$ we obtain the following effective action
\begin{equation}\label{particle_action}
 S = \int \dd t\, \left[- \vec{\bar r} \cdot  \partial_t \vec{k} - \epsilon_{\vec{k} - \vec A(\vec{\bar r})}\right]\,.
\end{equation} 

\begin{figure}
\begin{center}
\includegraphics[scale=0.5]{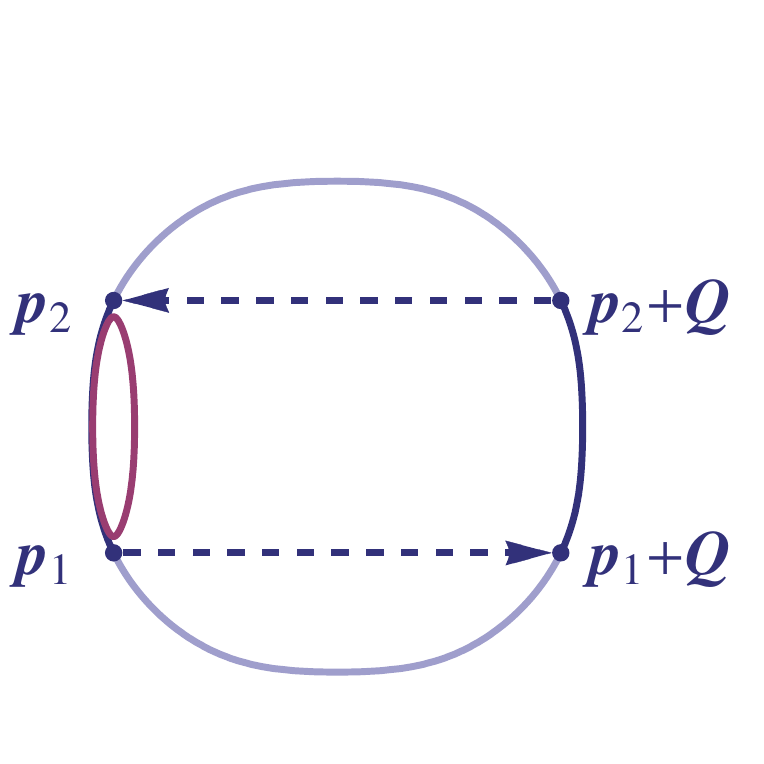}
\includegraphics[scale=0.5]{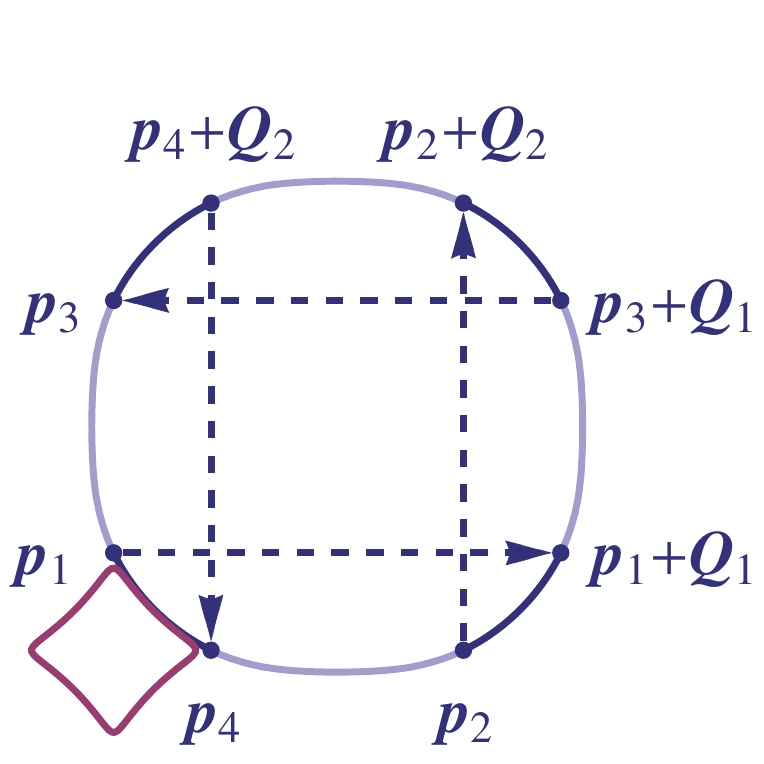}
\end{center}
\caption{\label{fig:semiclassical_plots} {\bf Fermi surface reconstruction in the presence of BDW order.} In blue, the Fermi surface, with points connected by a BDW wavevector $\vec Q$ in evidence. In red, the zeros of the reconstructed dispersion $E_{\vec p}$. On the right, reconstruction in the presence of two wavevectors $\vec Q_1$ and $\vec Q_2$.}
\end{figure}

This action describes the dynamics of the system as closely as possible while using the parametrization (\ref{eq:wavepacket}). The quantity $\epsilon_{\vec{k} - \vec A}$ is conserved on the equations of motion, as it plays the role of a Hamiltonian. The two components of $ \vec{p} \equiv \vec{k} - \vec A(\vec{\bar r})$ are canonically conjugate, and a complementary pair also exists: in the gauge $\vec A(\vec r) = B x \vec{\hat y}$ it is formed by $Y = \bar y - k_x / B$ and $k_y$. They satisfy
\begin{align}
 &\{p_x, p_y\} = B\,, \quad \{Y, k_y\} = 1\,, \\
 &\{p_i, Y\} = \{p_i, k_y\} = 0\,.
\end{align} 

The pair $(Y, k_y)$ decouples, and the equations of motion prescribe that $p_x$, $p_y$ describe an orbit $\epsilon_{\vec p} = \text{const}$. Carrying out Bohr-Sommerfeld quantization of the motion of this $(p_x, p_y)$ particle, we have the condition
\begin{equation}
 \frac{1}{B}\oint_{\epsilon_{\vec p} = \omega_n} p_x \dd p_y = 2\pi n \hbar\,,
\end{equation} 
that determines the energy $\omega_n$ of the Landau levels. Alternatively, there will be a peak in the density of states at the Fermi level when the magnetic field satisfies
\begin{equation}
 B_n = \frac{1}{2\pi n \hbar} \oint_{\epsilon_{\vec p} = 0} p_x \dd p_y\,.
\end{equation} 

Now we include a periodic density wave modulation in the hamiltonian:
\begin{align}\label{eq:single_particle_hamiltonian}
 &h = \sum_{\vec r, \vec a} \left[-t_{\vec a} + P_{\vec a} \cos[\vec Q \cdot (\vec r + \vec a / 2)]\right] e^{\ii \int \vec A\cdot \dd \vec l} \ket{\vec r + \vec a}\bra{\vec r}
\end{align}

Since the potential $P$ can scatter the momentum $\vec k$ to $\vec k + \vec Q$, it is necessary to consider a more general wavepacket
\begin{equation}
 \ket{\psi(t)} = \sum_{\vec r} \left[u(t) + v(t) e^{\ii \vec Q \cdot \vec r} \right]e^{\ii \vec{k}(t) \cdot \vec r -(\vec r - \vec{\bar r}(t))^2/2 L^2} \ket{\vec r}\,,
\end{equation} 
with real $u$, $v$.

Substituting in (\ref{eq:lattice_action}) and assuming that $L$ is much bigger than both the lattice spacing and the period of the density wave, we have
\begin{equation}
 S = \int \dd t\, \left[- \vec{\bar r} \cdot  \partial_t \vec{k} - 
\begin{pmatrix} u & v\end{pmatrix}
\begin{pmatrix} \epsilon_{\vec k - \vec A} & P_{\vec k + \vec Q / 2 - \vec A}\\ P_{\vec k + \vec Q / 2 - \vec A} & \epsilon_{\vec k + \vec Q - \vec A} \end{pmatrix}
\begin{pmatrix} u \\ v\end{pmatrix}\right]\,.
\end{equation} 

We take $u(t) = u_{\vec k(t) - \vec A(\vec{ \bar x}(t))}$, $v(t) = v_{\vec k(t) - \vec A(\vec{ \bar x}(t))}$, where $u_{\vec k}$, $v_{\vec k}$ satisfy the eigenvalue equation
\begin{equation}\label{eq:reconstruction_eigenvalues}
\begin{pmatrix} \epsilon_{\vec k} & P_{\vec k + \vec Q / 2}\\ P_{\vec k + \vec Q / 2} & \epsilon_{\vec k + \vec Q} \end{pmatrix}
\begin{pmatrix} u_{\vec k} \\ v_{\vec k}\end{pmatrix} = E_{\vec k}  \begin{pmatrix} u_{\vec k} \\ v_{\vec k}\end{pmatrix}\,,
\end{equation} 
and the action becomes identical to the case $P = 0$, only with the dispersion $\epsilon_{\vec{p}}$ changed to $E_{\vec{p}}$.

An intuitive picture of the dynamics of the system is given in Fig.~\ref{fig:semiclassical_plots}. As a first approximation, thinking of $P$ as a small perturbation, the electron starting at $\vec p_2$ moves along the Fermi surface more or less unperturbed until $\vec p_1$. Since $\vec p_1$ and $\vec p_1 + \vec Q$ are strongly mixed, the electron gets scattered by $P$ to $\vec p_1 + \vec Q$, then proceeds to $\vec p_2 + \vec Q$, and gets scattered back to $\vec p_2$, completing an orbit. Moving to the more accurate description above, the wavepacket travels along the curve $E_{\vec p} = 0$, the red pocket in the figure. On the left side of the pocket $u \simeq 1, v\simeq 0$, whereas $u \simeq 0, v\simeq 1$ on the right side, so the electron can be thought of having momentum $\vec p$ on one side, and $\vec p + \vec Q$ on the other. The scattering due to $P$ is then realized as the sharp transition in $u$, $v$ near $\vec p_1$ and $\vec p_2$.

Our discussion neglected higher order processes in $P$ that can scatter the electron from $\vec p$ to $\vec p + n \vec Q$, because these are suppressed for small $P$. In principle, however, they can be included by considering a more general wavepacket carrying a superposition of all relevant momenta. A more important generalization on the same line is necessary if two density waves with different wavevectors are present. For example, Fig.~\ref{fig:semiclassical_plots} also shows the case of two orthogonal wavevectors $\vec Q_1$ and $\vec Q_2$, which yield an orbit involving four patches of the Fermi surface as a lowest order process. In order to describe this orbit, a superposition of the four momenta $\vec k$, $\vec k + \vec Q_1$, $\vec k + \vec Q_2$, $\vec k + \vec Q_1 + \vec Q_2$ is necessary. Through an analysis very similar to the one above, we are led to the eigenvalue equation

\begin{equation}\label{eq:semiclassical_eigenvalue}
\begin{pmatrix} 
\epsilon_{\vec k} & P^{1}_{\vec k + \vec Q_1 / 2} & 0 & P^{2}_{\vec k + \vec Q_2/2}\\ 
P^{1}_{\vec k + \vec Q_1 / 2} & \epsilon_{\vec k + \vec Q_1} & P^{2}_{\vec k + \vec Q_1 + \vec Q_2/2} & 0\\
0 & P^{2}_{\vec k + \vec Q_1 + \vec Q_2 / 2} & \epsilon_{\vec k + \vec Q_1 + \vec Q_2} & P^{1}_{\vec k + \vec Q_1/2 + \vec Q_2}\\
P^{2}_{\vec k + \vec Q_2/2} & 0 & P^{1}_{\vec k + \vec Q_1 / 2 + \vec Q_2} & \epsilon_{\vec k + \vec Q_2}
\end{pmatrix}
\begin{pmatrix} u_{1} \\ u_{2}\\u_{3}\\u_{4}\end{pmatrix} = E_{\vec k}  \begin{pmatrix} u_{1} \\ u_{2}\\u_{3}\\u_{4}\end{pmatrix}\,,
\end{equation} 

analogous to (\ref{eq:reconstruction_eigenvalues}). The wavepacket travels along the curve $E_{\vec p} = 0$, shown in red in Fig.~\ref{fig:semiclassical_plots}. Additional scattering processes $\vec p\to \vec p + n_i \vec Q_i$ can also be included, but the computation becomes rather involved.

\subsection{Exact treatment of Quantum oscillations}
\label{QO:exact}

\begin{figure}
\begin{center}
\includegraphics[scale=0.78]{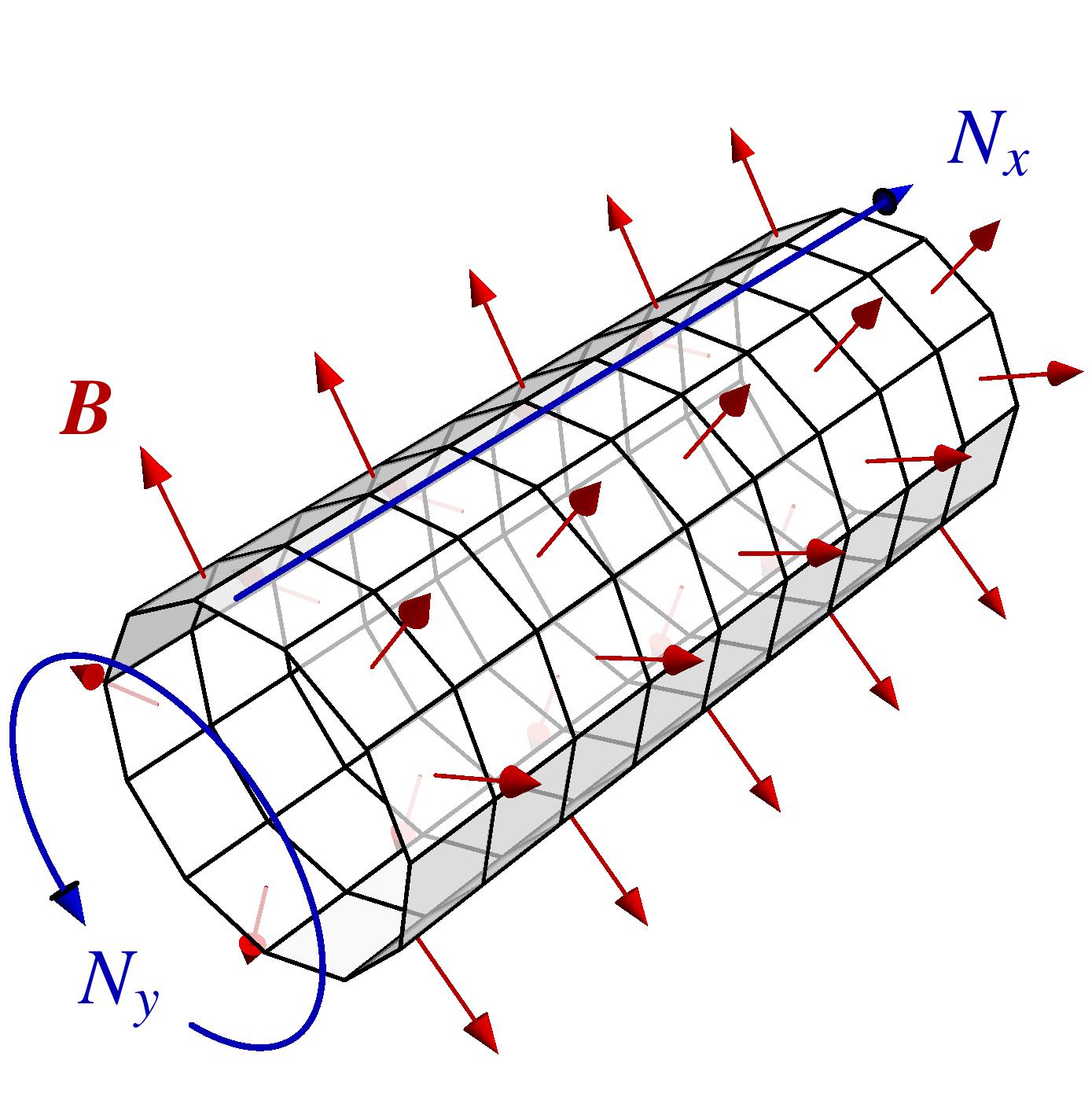}
\end{center}
\caption{{\bf Geometry of the setup.} Schematic diagram of the geometry used for the exact treatment of quantum oscillations.}
\label{fig:cylinder}
\end{figure}

Here we describe an exact treatment of both the density wave order
and the magnetic field acting on the full Schr\"odinger equation for the electrons.
The density of states at the Fermi level is given by
\begin{equation}
 D = \frac{1}{\pi} \mathrm{Im}\, \mathrm{Tr}\ \frac{1}{H - \ii \eta}\,,
\end{equation} 
where $H$ is given by (\ref{eq:single_particle_hamiltonian}). By ordering appropriately the two dimensional index $\vec r$, the matrix $H$ can be cast to block-tridiagonal form. For a lattice of $N_x \times N_y$ sites, the blocks have size $2N_y$ if second and third neighbor hopping is allowed, and there are $N_\tn{b} = N_x / 2$ of them.
\begin{equation}
 H = \begin{pmatrix}
     h_{11} & t_{12} & 0 & \ldots \\ 
     t_{21} & h_{22} & t_{23} & \ldots  \\
     0 & t_{32} & h_{33} & \ldots  \\
    \vdots & \vdots&\vdots
     \end{pmatrix}\,.
\end{equation} 

The diagonal blocks of $G = H^{-1}$, and hence the trace, can be efficiently ($\text{time} \sim N_x N_y^3$) calculated with the following iterative algorithm \cite{LeeFisher1981},
\begin{equation}
\begin{split}
&\begin{aligned}
 &L_1 = 0\,,
 &&
 L_{i+1} = t_{i+1,i}\left( h_{ii} - L_i \right)^{-1} t_{i, i+1}\\
 & R_{N_\tn{b}} = 0\,,
 &&
 R_{i-1} = t_{i-1,i}\left( h_{ii} - R_i \right)^{-1} t_{i, i-1}
\end{aligned}\\
&\;G_{ii} = \left( h_{ii} - L_i - R_i \right)^{-1}\,.
\end{split}
\end{equation} 
Since the computational cost scales only linearly in $N_x$, it is possible to take $N_x \sim 1000$, and finite size effect in this direction are negligible. Therefore, it is possible to have open boundary conditions as shown in Fig.~\ref{fig:cylinder}, and vary the magnetic field continuously, since the flux through the lattice need not be quantized. On the other hand, $N_y$ is constrained to be much smaller, $N_y \sim 50$, and it is necessary to impose periodic boundary conditions in this direction. As a consequence, the wavevectors of density wave modulations must satisfy the constraint $Q_y N_y =  2\pi n$, a mild commensuration constraint.

\subsection{Fermion self-energy}
\label{SF:self}

The naive Matsubara sum
\begin{equation}\label{H012}
H_0(\omega - \epsilon, \e)\equiv \frac{1}{\beta}\sum_n\frac{1}{\omega + i\Omega_n-\epsilon}\frac{1}{|\Omega_n|+\e}\,, 
\end{equation} 
although convergent, has poles for both $\Im \omega > 0$ and $\Im \omega < 0$. We are however free to add to $H_0$ the function
\begin{equation}
H_1(\omega, \epsilon,\e)\equiv\frac{n_\tn{F}(\epsilon)+n_\tn{B}(\epsilon-\omega)}{i(\epsilon-\omega)+\e}\,,
\end{equation} 
because $H_1(\ii \omega_n, \epsilon, \varepsilon) = 0$. It is easy to verify that the function
\begin{equation}
 H(\omega,\epsilon,\e) \equiv H_0(\omega - \epsilon,\e) + H_1(\omega,\epsilon,\e)
\end{equation}
is analytic for $\Im \omega > 0$, and hence yields the retarded self-energy. Note that $H_0$ is a real function of $\omega$ whereas $H_1$ has both a real and an imaginary part. Moreover, as a consequence of unitarity,
\begin{equation}
 \Im H(\omega,\epsilon,\e) = \Im H_1(\omega,\epsilon,\e) \leq 0\,.
\end{equation}
An explicit expression for $H_0$ exists in terms of the digamma function (see supplementary methods C). 

In terms of $H$ we have
\begin{equation}
\begin{split}
&\Sigma_\tn{s}(\vec k,\omega)=-\int_{\vec q}~\D^2_{\vec k-\frac{\vec q}{2}}\ H^{\star}(-\omega, \epsilon_{\vec q-\vec k},\e^\tn{s}_{\vec q})\,,\\
&\Sigma_\tn{b}(\vec k,\omega)=\sum_i\int_{\vec q}~P^{i\ 2}_{\vec k+\frac{\vec q}{2}}\ H(\omega,\epsilon_{\vec k+\vec q},\e^\tn{b}_{\vec q-\vec Q_i}).\
\end{split}
\end{equation} 
\\

{\bf Acknowledgements}
We thank E.~Altman, S.~Banerjee, E.~Berg, J.~C.~S.~Davis, P.~Johnson, 
M.~R.~Norman, C.~Proust, M.~Punk, S.~Sebastian, B.~Swingle, E.~D.~Torre and I.~Vishik for useful discussions. 
The research was supported by the U.S.\ National Science Foundation under grants DMR-1103860 and DMR-1360789, by the MURI grant W911NF-14-1-0003 from ARO, and by the Templeton Foundation. This research was also supported in part by Perimeter Institute for Theoretical Physics; research at Perimeter Institute is supported by the Government of Canada through Industry Canada and by the Province of Ontario through the Ministry of Research and Innovation.

{\bf Author contributions} A.A. conceived and performed the quantum oscillation computations. D.C. conceived and performed
the spectral function computations. All authors participated equally in sharpening the results and analysis, and in the writing of the manuscript.

%{\bf Competing financial interests} The authors declare no competing financial interests.

\begin{widetext}
\section{Supplementary Information}
\section*{Supplementary Figures}
\begin{figure}[h!]
\begin{center}
\includegraphics[scale=0.6]{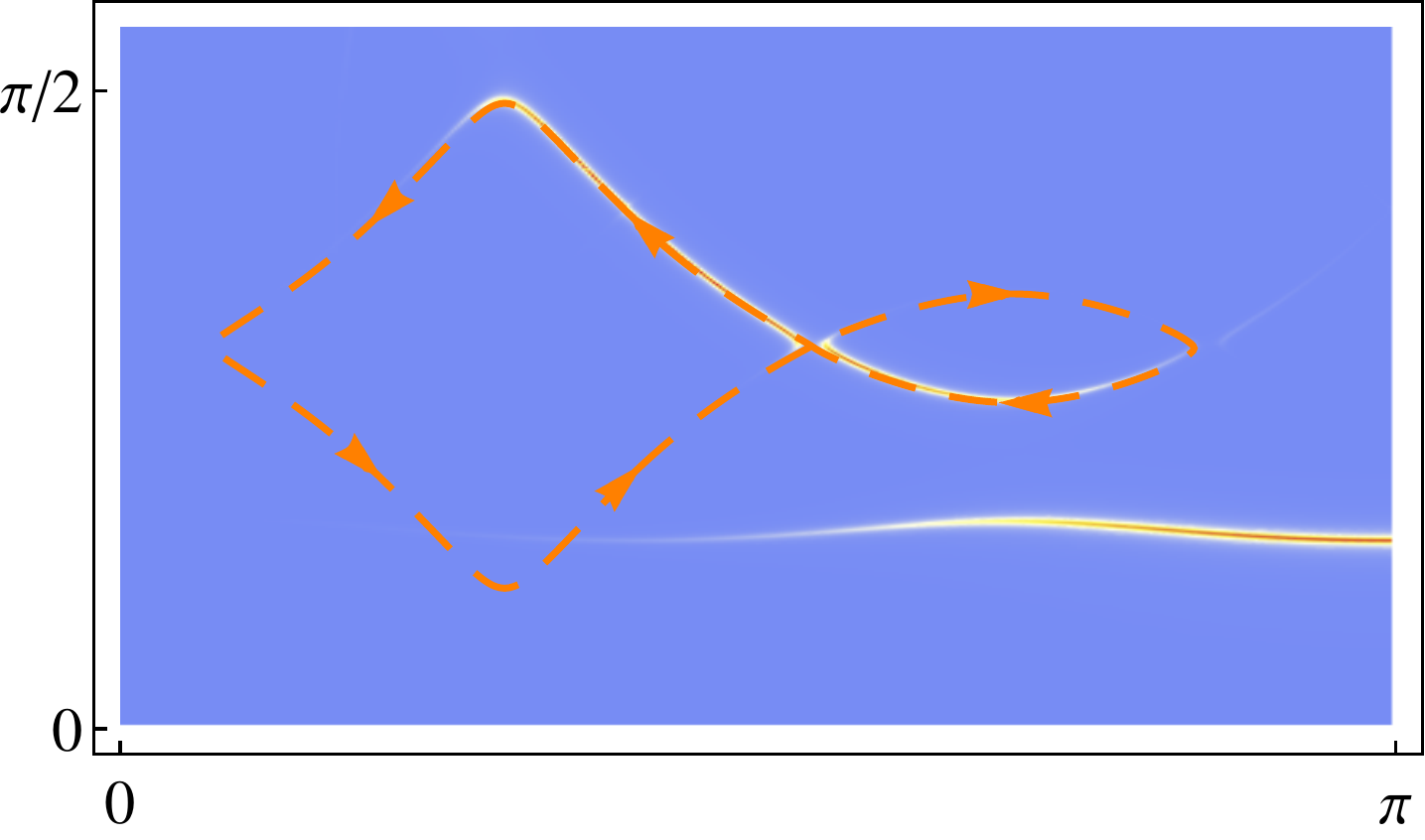}
\end{center}
%\rule{\linewidth}{1pt}
\caption{\label{fig:magnetic_breakdown}{\bf Magnetic breakdown.} Closeup of the electron spectral function covering the electron pocket and a hole pocket. In evidence a resonant orbit that involves tunneling through a small gap (magnetic breakdown). It is associated with the peaks in Fig.~{\color{blue}4} of the main article marked with golden bands. $p = 10\%$, $P^x_0 = 0.15$, $P^y_0/P^x_0 = 0.2$, $\delta = 0.3$.}
\end{figure}
\begin{figure}[h!]
\begin{center}
\includegraphics[scale=0.2]{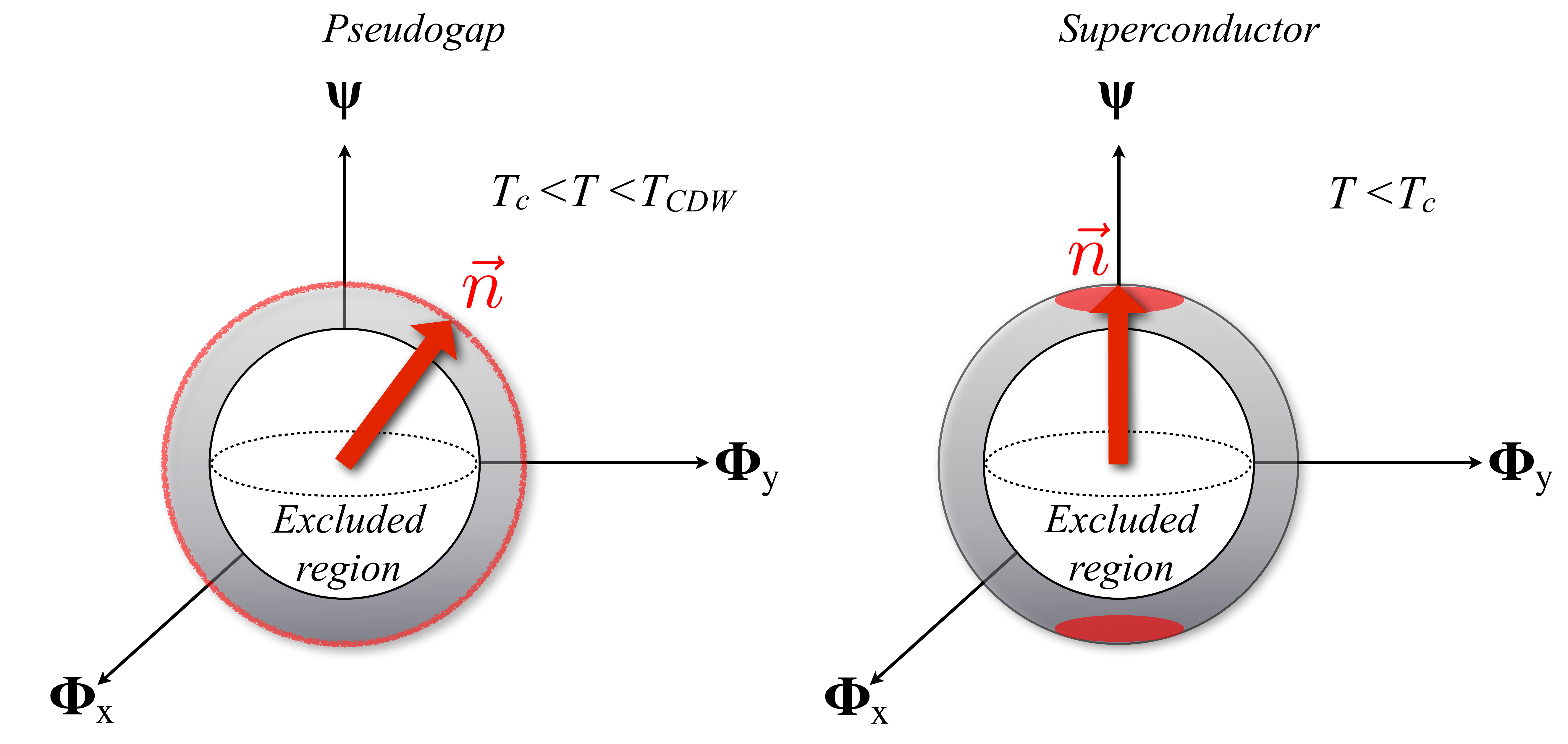}
\end{center}
\caption{{\bf O(6) Model.} Angular fluctuations of a multi-component order parameter $\vec{n}$ govern the pseudogap physics at zero magnetic field. For $T_\tn{c}<T<T_{\tn{CDW}}$, $\vec{n}$ fluctuates along all directions in phase space, whereas for $T<T_{\tn{c}}$, $\vec{n}$ preferentially fluctuates only along the superconducting axis.}
\label{fig:O(6)model}
\end{figure}
%%%%%%%%%%%%%%%%%%%%%%%%%%%%%%%%%%%%%%%%%%%%%%%%%%%%%

%%%%%%%%%%%%%%%%%%%%%%%%%%%%%%%%%%%%%%%%%%%%%%%%%%%%%%%%%%%
\begin{figure}[h!]
\begin{center}
\includegraphics[scale=0.5]{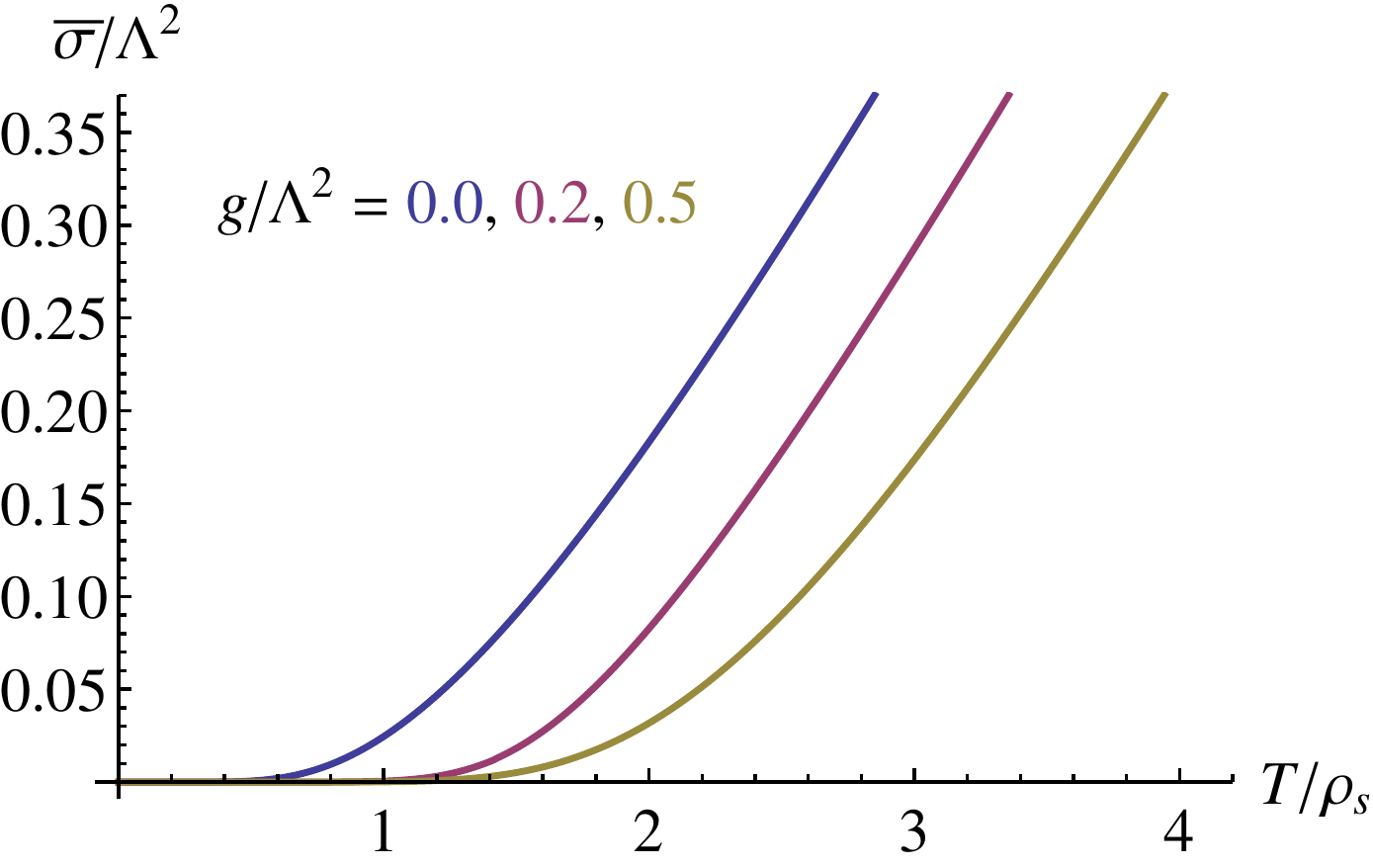}
\end{center}
\caption{{\bf Temperature dependence of $\ol\sigma/\Lambda^2$.} Obtained by solving (\ref{speqn}) with $\lambda=1.0$ and $w=0$ for $g/\Lambda^2 = 0,\, 0.2,\, 0.5$. }
\label{sigT}
\end{figure}
%%%%%%%%%%%%%%%%%%%%%%%%%%%%%%%%%%%%%%%%%%%%%%%%%%%%%%%%%%%

%%%%%%%%%%%%%%%%%%%%%%%%%%%%%%%%%%%%%%%%%%%%%%%%%%%%%%%%%%%
\begin{figure}[h!]
\begin{center}
\includegraphics[scale=0.3]{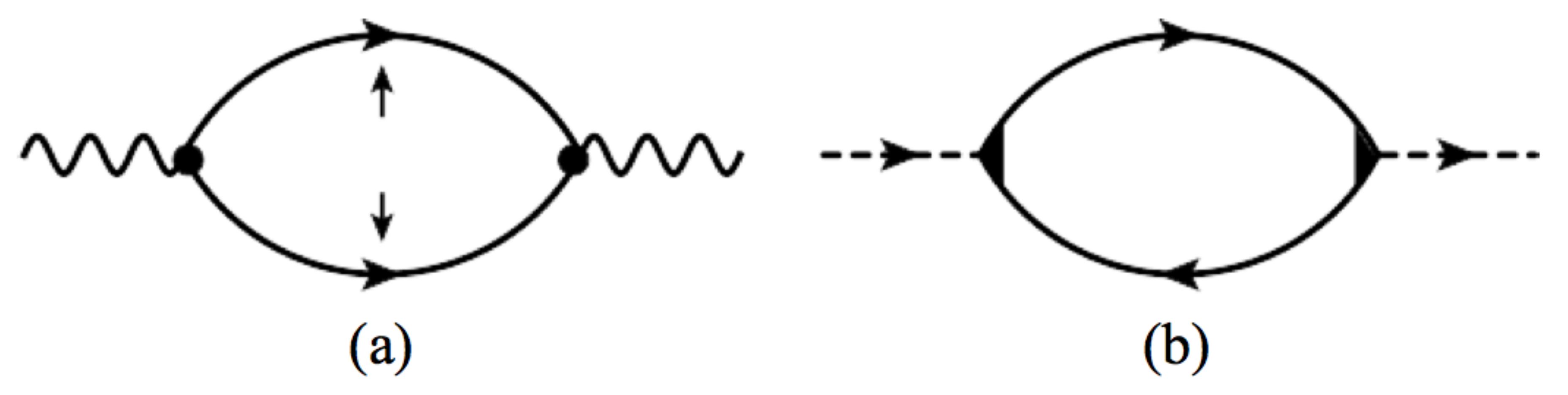}
\end{center}
\caption{{\bf Bosonic self-energies.} The one-loop self-energy contribution to the (a) Cooper pair ($\Pi^\tn{s}$) and (b) Bond order ($\Pi^\tn{b}$) propagators after integrating out the fermions.}
\label{selfphi}
\end{figure}
%%%%%%%%%%%%%%%%%%%%%%%%%%%%%%%%%%%%%%%%%%%%%%%%%%%%%%%%%%%

\newpage

\section*{Supplementary Notes 1}
\label{mb:qo}
{\bf Magnetic breakdown:} As discussed in the section on quantum oscillations, we observed an extra frequency in Fig. {\color{blue}4} of the main article, which did not correspond to the area of any pocket. It is only present when the amplitude on the second wavevector is relatively small, and hence the gap separating the elecron pocket from the blue hole pocket is really small. From a semiclassical point of view, it is then possible for the wavepacket coordinate $\vec p$ to tunnel through this gap and complete the orbit marked in supplementary fig.~\ref{fig:magnetic_breakdown} by the dashed line. Naively one would guess that the oscillation frequency due to this orbit should be given by the sum of the areas of the two pockets. However, the two pocket are encircled with opposite orientations, so the oscillation frequency is given instead by the difference of the two areas.  

%\subsection{Angular variation of spectral function}
%An alternative way of plotting the data shown in Fig.~{\color{blue} 9} is to show a color-plot of the spectral function as a function of frequency and the function $\cos(k_x)-\cos(k_y)$ (Fig.~\ref{angvar}), with $\vec k$ on the Fermi surface. The position of the tip of the arc, measured by the point at which the high intensity region along $\omega=0$ first splits up, and the maximum gap, measured by half of the maximum separation between the two branches, are clearly visible.

\section*{Supplementary methods}
\subsection{O(6) Model}
\label{SI:O6}
The O(6) model for fluctuating superconductivity and bond density wave, as introduced originally in ref.[{\color{blue} 1}] can be represented pictorially (see supplementary fig.~\ref{fig:O(6)model}). In this paper, we are interested in the regime where $\vec n$ fluctuates along all directions, as shown in the first figure.

In order to solve for the temperature dependence of $\ol\sigma$, the saddle point equations must be regulated, and we employ a hard momentum cutoff $\vec q^2 < \Lambda^2$. We have
\begin{equation}\label{speqn}
\begin{split}
&\frac{\rho_S}{T}=\frac{1}{4\pi}\left[\log\left(\frac{\Lambda^2+\ol\sigma}{\ol\sigma}\right) +  \frac{2}{\lambda}\log\left(\frac{\lambda\Lambda^2+\ol\sigma+g+\ol\phi}{\ol\sigma+g+\ol\phi} \right) \right],\\
&\ol\phi=\frac{wT}{2\pi\lambda\rho_S}\log\left(\frac{\lambda\Lambda^2+\ol\sigma+g+\ol\phi}{\ol\sigma+g+\ol\phi} \right). 
\end{split}
\end{equation} 

On dimensional grounds, the equations above are invariant under 
\begin{align}
& \Lambda \mapsto  b\, \Lambda && (g, \bar \sigma, \bar \phi) \mapsto b^2 (g, \bar \sigma, \bar \phi)\,,
\end{align}
and $\ol\phi$ is non-zero only if $w\neq0$.  The universal function $\sigma/\Lambda^2$ for a few choice parameters and $w = 0$ is shown in supplementary fig.~\ref{sigT}.

\subsection{Bosonic self-energy}
\label{BSE}
Let us compute the frequency dependence that arises due to the coupling of $\Psi$ and $\Phi$ to the underlying fermions. We want to evaluate the bubble diagrams of the type shown in supplementary fig.~\ref{selfphi}. 

We start by evaluating the diagram in supplementary fig.~\ref{selfphi}(a). It is given by,
\beq
\Pi^\tn{s}(\vec q,i\omega_n)=2\int_{\vec k}\D_{\vec k}^2\frac{1-n_F(\epsilon_{\vec k+\vec q/2,\uparrow})-n_F(\epsilon_{-\vec k+\vec q/2,\downarrow})}{i\omega_n-\epsilon_{\vec k+\vec q/2,\uparrow}-\epsilon_{-\vec k+\vec q/2,\downarrow}},
\eeq
where $i\omega_n$ is a bosonic Matsubara frequency.
The leading contribution to this diagram comes from the quasiparticles in the vicinity of the antinodes, where $|\D_{\vec k}|\approx2\D_0$ is maximum. Let us therefore analyze the behavior of $\Pi^\tn{s}$ at $\vec q=0$ by expanding in the vicinity of the antinodes. We can expand $\epsilon_{\vec k}=\vec v.\vec k$ where the Fermi velocities at the antipodal points are related by symmetry [{\color{blue}2}].   The imaginary part of the retarded bubble at real frequencies, $\omega$ ($i\omega_n\rightarrow \omega+i0^+$), then evaluates to,
\beq
\tn{Im}\Pi^\tn{s}_\tn{R}(\vec q=0,\omega)=\frac{8(2\D_0)^2}{4\pi v_xv_y}\int dxdy[1-2n_\tn{F}(x+y)]\delta(\omega-2(x+y)),
\eeq
where $\vec v^2=v_x^2+v_y^2$ with $v_y>v_x$ and we have considered contributions from all the antinodal patches. By restricting ourselves to the region $|\vec k|<\Lambda$ and in the small frequency regime ($\omega<T\ll\vec v\Lambda$), 
\beq
\tn{Im}\Pi^\tn{s}_\tn{R}(\vec q=0,\omega)=\frac{32\D_0^2\Lambda}{4\pi v_y}\tanh\bigg(\frac{\omega}{4T}\bigg)\approx\frac{2\D_0\Lambda}{\pi v_y}~\omega,
\eeq 
where we have used the fact that at low temperatures, the $1/T$ behavior gets cutoff by $\D_0$. The linear dependence on $\omega$ implies a Landau damped form in the propagator, which gives rise to many of the interesting features at low energies.

Similarly, we can compute the diagram in supplementary figure~\ref{selfphi} (b). In the low energy limit, we would like to evaluate the bubble due to the fermions in the antinodal regions that are nested by the CDW wavevector, $\vec Q_i$. The Fermi velocities at two such points in the vicinity of $(\pi,0)$ are given by $\vec v_1=(v_x,v_y)$ and $\vec v_2=(v_x,-v_y)$. The expression for the particle-hole bubble is given by the standard Lindhard-type form, 
\beq
\Pi^{\tn{b}}(\vec q,i\omega_n)=-2\int_{\vec k}P_{\vec k}^{i2}\frac{n_\tn{F}(\epsilon_{\vec k-\vec q/2}|_1)-n_\tn{F}(\epsilon_{\vec k+\vec q/2,2}|_2)}{i\omega_n+\epsilon_{\vec k-\vec q/2}|_1-\epsilon_{\vec k+\vec q/2}|_2},
\eeq
where $\epsilon_{\vec k}|_1=v_xk_x+v_yk_y$ and, $\epsilon_{\vec k}|_2=v_1k_x-v_2k_y$. In the antinodal region, $|P_{\vec k}^i|\approx 2P_0^i$ and $\omega_n$ is a bosonic Matsubara frequency. The imaginary part of the retarded bubble, Im$\Pi^\tn{b}_\tn{R}(\vec q=0,\omega)$ at a real frequency $\omega$ and evaluated at $T=0$ is given by,
\beq
\tn{Im}\Pi^\tn{b}_\tn{R}(\vec q=0,\omega)=-\frac{2(2P_0^i)^2}{4\pi^2v_xv_y}\int dx dy~[\theta(-x-y)-\theta(y-x)]\delta(\omega+2y),
\eeq
where $\theta(...)$ represents the Heaviside step function. 

In the limit of small frequencies, $\omega< v_x\Lambda\ll v_y\Lambda$, the above evaluates to,
\beq
\tn{Im}\Pi^\tn{b}_\tn{R}(\vec q=0,\omega)=\frac{(2P_0^i)^2}{4\pi^2v_xv_y}~\omega,
\eeq
which is once again a sign of the BDW propagator having a Landau damping term at low energies. 

\subsection{Real and imaginary part of Fermion self-energy}
\label{re}
It is possible to evaluate the sum over $n$ in eqn.29 of the main article analytically. We obtain
\begin{equation}
 H_0(\omega, \e) = -\frac{T}{\omega \e} - \frac{\omega\, \psi\left( \frac{\e}{2\pi T}\right)}{\pi(\e^2 + \omega^2)}  + \Re\left[\frac{\psi\left(\frac{\ii \omega}{2\pi T}\right)}{\pi (\omega + \ii \e)}\right]\,,
\end{equation} 
where $\psi(z)\equiv\Gamma'(z)/\Gamma(z)$ is the digamma function.

The real part and imaginary parts of the self energies have the explicit expression
\beq
\tn{Re}\Sigma_\tn{s}^\tn{R}(\vec k,\omega)&=&\int_{\vec q}\D_{\vec k-\frac{\vec q}{2}}^2\bigg[ H_0(-\omega-\epsilon_{-\vec k+\vec q},\epsilon_\tn{s}(\vec q))
+ [n_\tn{F}(\epsilon_{-\vec k+\vec q}) + n_\tn{B}(\epsilon_{-\vec k+\vec q}+\omega)]~\frac{\e_\tn{s}(\vec q)}{(\omega+\epsilon_{-\vec k + \vec q})^2+(\e_\tn{s}(\vec q))^2}\bigg]\nonumber\\
\tn{Re}\Sigma_\tn{b}^\tn{R}(\vec k,\omega)&=&\int_{\vec q}P_{\vec k+\frac{\vec q}{2}}^{i2}\bigg[ H_0(\omega-\epsilon_{\vec k+\vec q},\epsilon_\tn{b}(\vec q))
+ [n_\tn{F}(\epsilon_{\vec k+\vec q}) + n_\tn{B}(\epsilon_{\vec k+\vec q}-\omega)]~\frac{\e_\tn{b}(\vec q)}{(\omega-\epsilon_{\vec k + \vec q})^2+(\e_\tn{b}(\vec q))^2}\bigg],
\eeq
\beq
&&\tn{Im}\Sigma^\tn{R}_\tn{s}(\vec k,\omega)=\int_{\vec q}~\D^2_{\vec k-\frac{\vec q}{2}}~ [n_\tn{F}(\epsilon_{-\vec k+\vec q})+n_\tn{B}(\epsilon_{-\vec k+\vec q}+\omega)]~\frac{\omega+\epsilon_{-\vec k+\vec q}}{(\omega+\epsilon_{-\vec k+\vec q})^2+(\e_\tn{s}(\vec q))^2},\\
&&\tn{Im}\Sigma^\tn{R}_\tn{b}(\vec k,\omega)=\int_{\vec q}~P^{i2}_{\vec k+\frac{\vec q}{2}}~ [n_\tn{F}(\epsilon_{\vec k+\vec q})+n_\tn{B}(\epsilon_{\vec q+\vec k}-\omega)]~\frac{\omega-\epsilon_{\vec k+\vec q}}{(\omega-\epsilon_{\vec k+\vec q})^2+(\e_\tn{b}(\vec q))^2}.
\eeq
\section*{Supplementary References}
\begin{enumerate}
\item L. Hayward, D. Hawthorn, R. Melko and S. Sachdev, Angular Fluctuations of a Multicomponent Order Describe the Pseudogap of YBa$_2$Cu$_3$O$_{6+x}$, Science {\bf 343}, 1336-1339 (2014).

\item D. Chowdhury and S. Sachdev, Feedback of superconducting fluctuations on charge order in the underdoped cuprates, Phys. Rev. B {\bf 90}, 134516 (2014). 
\end{enumerate}
\end{widetext}


\begin{thebibliography}{99}

\bibitem{alloul} H.~Alloul, P.~Mendels, G.~Collin, and P.~Monod, 
$^{89}$Y NMR Study of the Pauli Susceptibility of the CuO$_2$ Planes in YBa$_2$Cu$_3$O$_{6+x}$,
Phys. Rev. Lett. {\bf 61}, 746 (1988).

\bibitem{YRZ} Kai-Yu Yang, T. M. Rice and Fu-Chun Zhang, Phenomenological theory of the pseudogap state, Phys. Rev. B
  {\bf 73}, 174501 (2006).

\bibitem{Wen12} Jia-Wei Mei, Shinji Kawasaki, Guo-Qing Zheng, Zheng-Yu Weng, and Xiao-Gang Wen, Luttinger-volume violating Fermi liquid in the pseudogap phase of the cuprate superconductors, Phys. Rev. B {\bf 85}, 134519 (2012).

\bibitem{Balents05} L. Balents, L. Bartosch, A. Burkov, S. Sachdev, and K. Sengupta, 
Putting competing orders in their place near the Mott transition,  Phys. Rev. B {\bf 71}, 144508 (2005).

\bibitem{PS12} M. Punk and S. Sachdev, Fermi surface reconstruction in hole-doped $t$-$J$ models without long-range antiferromagnetic order, Phys. Rev. B {\bf 85}, 195123 (2012).

\bibitem{DCSS14b} D. Chowdhury and S. Sachdev, Density wave instabilities of fractionalized Fermi liquids, Preprint at http://arxiv.org/abs/1409.5430 (2014).

\bibitem{Johnson11} H.-B. Yang {\it et al. \/}, Reconstructed Fermi Surface of Underdoped Bi$_2$Sr$_2$CaCu$_2$O$_{8+\delta}$ Cuprate Superconductors, Phys. Rev. Lett. {\bf 107}, 047003 (2011).
 
\bibitem{KivelsonRMP} S. A. Kivelson, I. P. Bindloss, E. Fradkin, V. Oganesyan, J. M. Tranquada, A. Kapitulnik, and C. Howald, How to detect fluctuating stripes in the high-temperature superconductors, 
Rev. Mod. Phys. {\bf 75}, 1201 (2003).

\bibitem{SSRMP} S.~Sachdev,  Order and quantum phase transitions in the cuprate superconductors, Rev. Mod. Phys. {\bf 75}, 913 (2003).

\bibitem{pines} J.~Schmalian, D.~Pines, and B.~Stojkovi\'c, Weak Pseudogap Behavior in the Underdoped Cuprate Superconductors, Phys. Rev. Lett. {\bf 80}, 3839 (1998).

\bibitem{kampf} A.P. Kampf and J.R. Schrieffer, Spectral function and photoemission spectra in antiferromagnetically correlated metals, Phys. Rev. B {\bf 42}, 7967 (1990).

\bibitem{so5} S.-C.~Zhang, 
``A Unified Theory Based on SO(5) Symmetry of Superconductivity and Antiferromagnetism,''
Science {\bf 275}, 1089 (1997).

\bibitem{SSAF13} L.~E.~Hayward, D.~G.~Hawthorn, R.~G.~Melko and S. Sachdev, Angular Fluctuations of a Multicomponent Order Describe the Pseudogap of YBa$_2$Cu$_3$O$_{6+x}$, Science {\bf 343}, 1336-1339 (2014).

\bibitem{MHJ11} T. Wu {\it et al. \/}, Magnetic-field-induced charge-stripe order in the high-temperature superconductor YBa$_2$Cu$_3$O$_y$, Nature {\bf 477}, 191-194 (2011).

\bibitem{Ghi12} G. Ghiringelli {\it et al.\/}, Long-Range Incommensurate Charge Fluctuations in (Y,Nd)Ba$_2$Cu$_3$O$_{6+x}$, Science {\bf 337}, 821-825 (2012).

\bibitem{DGH12} A.J. Achkar {\it et al.\/}, Distinct Charge Orders in the Planes and Chains of Ortho-III-Ordered YBa$_2$Cu$_3$O$_{6+\delta}$ Superconductors Identified by Resonant Elastic X-ray Scattering, Phys. Rev. Lett. {\bf 109}, 167001 (2012).

\bibitem{SH12} J. Chang {\it et al.\/}, Direct observation of competition between superconductivity and charge density wave order in YBa$_2$Cu$_3$O$_{6.67}$, Nat. Phys. {\bf 8}, 871-876 (2012).

\bibitem{MHJ13} T. Wu {\it et al.\/}, Emergence of charge order from the vortex state of a high-temperature superconductor, Nat. Comms. {\bf 4}, 2113 (2013). 

\bibitem{SSAF14} L. E. Hayward, A.~J.~Achkar, D.~G~Hawthorn, R.~G.~Melko and S. Sachdev, 
Diamagnetism and density wave order in the pseudogap regime of YBa$_2$Cu$_3$O$_{6+x}$, 
Phys. Rev. B {\bf 90}, 094515 (2014)

\bibitem{LT07} N. Doiron-Leyraud {\it et al.\/}, Quantum oscillations and the Fermi surface in an underdoped high-$T_\tn{c}$ superconductor, 
Nature {\bf 447}, 565-568 (2007). 

\bibitem{leboeuf} D. LeBoeuf {\it et al.\/}, Electron pockets in the Fermi surface of hole-doped high-T$_\tn{c}$ superconductors, Nature {\bf 450}, 533-536 (2007)

\bibitem{louis2} L.~Taillefer, Fermi surface reconstruction in high-$T_{\tn{c}}$ superconductors
J.~Phys.: Condens. Matter {\bf 21}, 164212 (2009).

\bibitem{harrison} N. Harrison and S. E. Sebastian, Protected Nodal Electron Pocket from Multiple-Q Ordering in Underdoped High Temperature Superconductors, Phys. Rev. Lett. {\bf 106}, 226402 (2011).

\bibitem{suchitra2} S.~E.~Sebastian, N.~Harrison and G.~G.~Lonzarich, Towards resolution of the Fermi surface in underdoped high-$T_\tn{c}$ superconductors, Rep. Prog. Phys. {\bf 75}, 102501 (2012).

\bibitem{greven} N.~Bari\u{s}i\'{c} {\it et al.\/}, Universal quantum oscillations in the underdoped cuprate superconductors, Nature Physics {\bf 9}, 761-764 (2013).

\bibitem{suchitra3} S.~E.~Sebastian {\it et al.\/}, Normal-state nodal electronic structure in underdoped high-$T_{\tn{c}}$ copper oxides, Nature {\bf 511}, 61-64 (2014).

\bibitem{ZXS03} A. Damascelli, Z. Hussain and Z.-X. Shen, Angle-resolved photoemission studies of the cuprate superconductors, Rev. Mod. Phys. {\bf 75}, 473 (2003).

\bibitem{kanigel06} A. Kanigel {\it et al.\/}, Evolution of the pseudogap from Fermi arcs to the nodal liquid, Nat. Phys. {\bf 2}, 447-451 (2006).

\bibitem{PJ08} H. B. Yang {\it et al.\/}, Emergence of preformed Cooper pairs from the doped Mott insulating state in Bi$_2$Sr$_2$CaCu$_2$O$_{8+\delta}$, Nature {\bf 456}, 77-80 (2008).

\bibitem{Hashimoto12} M. Hashimoto {\it et al.\/}, Particle-hole symmetry breaking in the pseudogap state of Bi2201, Nat. Phys. {\bf 6}, 414-418 (2010).

\bibitem{Inna12} I.M. Vishik {\it et al.\/}, Phase competition in trisected superconducting dome, Proc. Natl. Acad. Sci. {\bf 109}, 18332-18337 (2012). 

\bibitem{Kaminski13} T. Kondo {\it et al.\/}, Formation of Gapless Fermi Arcs and Fingerprints of Order in the Pseudogap State of Cuprate Superconductors, Phys. Rev. Lett. {\bf 111}, 157003 (2013).

\bibitem{Dessau12} T.J. Reber {\it et al.\/}, The origin and non-quasiparticle nature of Fermi arcs in Bi$_2$Sr$_2$CaCu$_2$O$_{8+\delta}$, Nat. Phys. {\bf 8}, 606-610 (2012).

\bibitem{MMSS10} M.~A.~Metlitski and S. Sachdev, Quantum phase transitions of metals in two spatial dimensions. II. Spin density wave order, Phys. Rev. B {\bf 82}, 075128 (2010).

\bibitem{SSRP13} S. Sachdev and R.L. Placa, Bond Order in Two-Dimensional Metals with Antiferromagnetic Exchange Interactions, Phys. Rev. Lett. {\bf 111}, 027202 (2013).

\bibitem{Efetov12} K. Efetov, H. Meier and C. Pepin, Pseudogap state near a quantum critical point, Nat. Phys. {\bf 9}, 442-446 (2012). 

\bibitem{Metzner1} T.~Holder and W.~Metzner, Incommensurate nematic fluctuations in two-dimensional metals, Phys. Rev. B {\bf 85}, 165130 (2012).

\bibitem{Metzner2} C.~Husemann and W.~Metzner, Incommensurate nematic fluctuations in the two-dimensional Hubbard model, Phys. Rev. B {\bf 86}, 085113 (2012).

\bibitem{Yamase} M.~Bejas, A.~Greco, and H.~Yamase, Possible charge instabilities in two-dimensional doped Mott insulators, Phys. 
Rev. B {\bf 86}, 224509 (2012).

\bibitem{Kee} Hae-Young Kee, C.~M.~Puetter, and D.~Stroud, Transport signatures of electronic-nematic stripe phases, J. Phys.: Condens. Matter {\bf 25}, 202201 (2013).

\bibitem{Kampf} S.~Bulut, W.~A.~Atkinson, and A.~P.~Kampf, Charge Order in the Pseudogap Phase of Cuprate Superconductors, Preprint at http://arxiv.org/abs/1404.1335 (2014).

\bibitem{DHL13} J.~C.~S\'eamus Davis and Dung-Hai Lee, Concepts relating magnetic interactions, intertwined electronic orders, and strongly correlated superconductivity, Proc. Natl. Acad. Sci. {\bf 110}, 17623-17630 (2013).

\bibitem{JSSS14} J. D. Sau and S. Sachdev, Mean-field theory of competing orders in metals with antiferromagnetic exchange interactions, Phys. Rev. B {\bf 89},  075129 (2014).

\bibitem{AASS14a} A. Allais, J. Bauer and S. Sachdev, Density wave instabilities in a correlated two-dimensional metal, Phys. Rev. B {\bf 90}, 155114  (2014).

\bibitem{AASS14b} A. Allais, J. Bauer and S. Sachdev, Auxiliary-boson and DMFT studies of bond ordering instabilities of t-J-V models on the square lattice, Ind. J of Phys., {\bf 88}, 9, pp 905 (2014).

\bibitem{KE13} H. Meier, C. Pepin, M. Einenkel and K. Efetov, Cascade of phase transitions in the vicinity of a quantum critical point, Phys. Rev. B {\bf 89}, 195115 (2014).

\bibitem{comin2} R.~Comin {\it et al.\/},The symmetry of charge order in cuprates, Preprint at http://arxiv.org/abs/1402.5415 (2014).

\bibitem{SSJSD14} K.~Fujita, M.~H.~Hamidian {\it et al.\/}, Direct phase-sensitive identification of a $d$-form factor density wave in underdoped cuprates, Proc. Natl. Acad. Sci. {\bf 111}, E3026 (2014).

\bibitem{DCSS14} D. Chowdhury and S. Sachdev, Feedback of superconducting fluctuations on charge order in the underdoped cuprates, Phys. Rev. B {\bf 90}, 134516 (2014). 

\bibitem{Hossain2008} M.~Hossain {\it et al.\/}, In situ doping control of the surface of high-temperature superconductors, Nature Phys. {\bf 4}, 527 (2008).

\bibitem{suchitra4} S.~E.~Sebastian {\it et al.\/}, Quantum oscillations from nodal bilayer magnetic breakdown in the underdoped high temperature superconductor YBa$_2$Cu$_3$O$_{6+x}$, Phys. Rev. Lett. {\bf 108}, 19 (2012).

\bibitem{vignolle} B. Vignolle, D. Vignolles, M. Julien, C. Proust, From quantum oscillations to charge order in high-$T_\tn{c}$ copper oxides in high magnetic fields, C. R. Physique {\bf 14}, 39-52 (2013).

\bibitem{blackburn} E. Blackburn, {\it et al.\/}, X-Ray Diffraction Observations of a Charge-Density-Wave Order in Superconducting Ortho-II YBa$_2$Cu$_3$O$_{6.54}$ Single Crystals in Zero Magnetic Field, Phys. Rev. Lett. {\bf 110}, 137004 (2013).

\bibitem{YZAMSK14} Y. Zhang, A.V. Maharaj and S. Kivelson, Are there quantum oscillations in an incommensurate charge density wave?, Preprint at http://arxiv.org/abs/1410.5108 (2014).

\bibitem{cyril} N. Doiron-Leyraud \etal, Evidence for a small hole pocket in the Fermi surface of underdoped YBa$_2$Cu$_3$O$_y$, Preprint at http://arxiv.org/abs/1409.2788 (2014).

\bibitem{nematic1} Y.~Kohsaka {\it et al.\/}, An Intrinsic Bond-Centered Electronic Glass with Unidirectional Domains in Underdoped Cuprates, Science {\bf 315}, 1380-1385 (2007).

\bibitem{nematic2} M.~J.~Lawler {\it et al.\/}, Intra-unit-cell electronic nematicity of the high-Tc copper-oxide pseudogap states, Nature {\bf 466}, 347-351 (2010).

\bibitem{MN07} A.~J.~Millis and M.~R.~Norman, Antiphase stripe order as the origin of electron pockets observed in 1/8-hole-doped cuprates, Phys. Rev. B {\bf 76}, 220503 (2007).

\bibitem{YLK11} Hong Yao, Dung-Hai Lee, and S.~Kivelson, Fermi-surface reconstruction in a smectic phase of a high-temperature superconductor, Phys. Rev. B {\bf 84}, 012507 (2011).

\bibitem{EWC12} Jonghyoun Eun, Zhiqiang Wang, and S.~Chakravarty, Quantum oscillations in YBa$_2$Cu$_3$O$_{6+\delta}$ from period-8 d-density wave order, Proc. Natl. Acad. Sci. {\bf 109}, 13198-13203 (2012).
 
\bibitem{Fradkin10} A.~Jaefari, S.~Lal, and E.~Fradkin, Charge-density wave and superconductor competition in stripe phases of high-temperature superconductors, Phys. Rev. B {\bf 82}, 144531 (2010).

\bibitem{EBEA07} E. Berg and E. Altman, Evolution of the Fermi Surface of d-Wave Superconductors in the Presence of Thermal Phase Fluctuations, Phys. Rev. Lett. {\bf 99}, 247001 (2007).

\bibitem{TSPAL09} T. Senthil and P.A. Lee, Synthesis of the phenomenology of the underdoped cuprates, Phys. Rev. B {\bf 79}, 245116 (2009).

\bibitem{SBTVR11} S. Banerjee, T.V. Ramakrishnan and C. Dasgupta, Effect of pairing fluctuations on low-energy electronic spectra in cuprate superconductors, Phys. Rev. B {\bf 84}, 144525 (2011).

\bibitem{SBMR13} S. Banerjee, S. Zhang and M. Randeria, Theory of quantum oscillations in the vortex-liquid state of high-$T_{\tn{c}}$ superconductors, Nat. Comms. {\bf 4}, 1700 (2013).

\bibitem{norman} T. Micklitz and M.R. Norman, Nature of spectral gaps due to pair formation in superconductors, Phys. Rev. B {\bf 80}, 220513(R) (2009).

\bibitem{Chubukov14} Y. Wang and A. V. Chubukov, Charge-density-wave order with momentum (2Q,0) and (0,2Q) within the spin-fermion model: continuous and discrete symmetry breaking, preemptive composite order, and relation to pseudogap in hole-doped cuprates, Phys. Rev. B {\bf 90}, 035149 (2014).

\bibitem{JH02} J. E. Hoffman {\it et al.\/}, A Four Unit Cell Periodic Pattern of Quasi-Particle States Surrounding Vortex Cores in Bi$_2$Sr$_2$CaCu$_2$O$_{8+\delta}$, Science {\bf 295}, 466-469 (2002).

\bibitem{aharon} C.~Howald, H.~Eisaki, N.~Kaneko, M.~Greven, and A.~Kapitulnik, Periodic density-of-states modulations in superconducting Bi$_2$Sr$_2$CaCu$_2$O$_{8+\delta}$, Phys. Rev. B {\bf 67}, 014533 (2003).

\bibitem{AY04} M. Vershinin, S. Misra, S. Ono, Y. Abe, Yoichi Ando, and A.~Yazdani, Local Ordering in the Pseudogap State of the High-Tc Superconductor Bi$_2$Sr$_2$CaCu$_2$O$_{8+\delta}$, Science {\bf 303}, 1995-1998 (2004).

\bibitem{comin13} R. Comin {\it et al.\/}, Charge Order Driven by Fermi-Arc Instability in Bi$_2$Sr$_{2-x}$La$_x$CuO$_{6+\delta}$, Science {\bf 343}, 390-392 (2014).

\bibitem{neto13} E.H. da Silva Neto {\it et al.\/}, Ubiquitous Interplay Between Charge Ordering and High-Temperature Superconductivity in Cuprates, Science {\bf 343}, 393-396 (2014).

\bibitem{JH14} Y.~He {\it et al.\/}, Fermi Surface and Pseudogap Evolution in a Cuprate Superconductor, Science {\bf 344}, 608-611 (2014). 

\bibitem{JSD14} K.~Fujita {\it et al.\/}, Simultaneous Transitions in Cuprate Momentum-Space Topology and Electronic Symmetry Breaking, Science {\bf 344}, 612-616 (2014). 

\bibitem{Ramshaw14} B.~J.~Ramshaw {\it et al.\/}, A quantum critical point at the heart of high temperature superconductivity,
Preprint at http://arxiv.org/abs/1409.3990 (2014).

 
\bibitem{LeeFisher1981} P.~A.~Lee and D.~S.~Fisher, Anderson Localization in Two Dimensions, Phys. Rev. Lett. {\bf 47} 882-885 (1981).



%\bibitem{DGH13} A.J. Achkar et al., Phys. Rev. Lett. {\bf 110}, 017001 (2013).

%\bibitem{CP13} D. LeBeouf {\it et al.\/}, Nature Phys. {\bf 9}, 79 (2013).

\end{thebibliography}
\end{document}